\documentclass[aps,tightenlines,twocolumn,floats,prd,nofootinbib,
superscriptaddress,showpacs,preprintnumbers]{revtex4-1}
\def\bea{\begin{eqnarray}}
\def\be{\begin{equation}}
\def\ee{\end{equation}}
\def\eea{\end{eqnarray}}
\def\bal{\begin{align}}
\def\eal{\end{align}}

\def\sfrac#1#2{{\textstyle \frac{#1}{#2}}}

\usepackage[usenames,dvipsnames]{color}
\usepackage{graphics}
\usepackage{graphicx}
\usepackage{epsf} 
\usepackage{amsmath}
\usepackage{amssymb}
\usepackage{bbold}
\usepackage{bm}
\usepackage{slashed}
\usepackage{float}
\usepackage{psfrag}
\usepackage{footnote} 
\usepackage{xcolor}
\usepackage{textcomp}
\def\peq #1{\scriptstyle{#1}}
\begin{document}

\title{Chiral symmetry and $\pi$-$\pi$ scattering in the Covariant Spectator Theory}



\author{Elmar P. Biernat}
\email[]{elmar.biernat@tecnico.ulisboa.pt}
\affiliation{Centro de F\'isica Te\'orica de Part\'iculas (CFTP), Instituto Superior T\'ecnico (IST), Universidade de Lisboa, 1049-001 Lisboa, Portugal}
\author{ M. T. Pe\~na}
\email[]{teresa.pena@tecnico.ulisboa.pt}

\affiliation{Centro de F\'isica Te\'orica de Part\'iculas (CFTP), Instituto Superior T\'ecnico (IST), Universidade de Lisboa, 1049-001 Lisboa, Portugal}
\affiliation{ Departamento de F\' isica, Instituto Superior T\'ecnico (IST), Universidade de Lisboa, 1049-001 Lisboa, Portugal}

\author{J. E. Ribeiro}
\email[]{emilioribeiro@tecnico.ulisboa.pt}

\affiliation{Centro de F\'isica das Intera\c c\~oes Fundamentais (CFIF), Instituto Superior T\'ecnico (IST), Universidade de Lisboa, 1049-001 Lisboa, Portugal}

\author{Alfred Stadler}
\email[]{stadler@uevora.pt}

\affiliation{Departamento de F\'isica, Universidade de 
\'Evora, 7000-671 \'Evora, Portugal} 

\affiliation{Centro de F\'isica Te\'orica de Part\'iculas (CFTP), Instituto Superior T\'ecnico (IST), Universidade de Lisboa, 1049-001 Lisboa, Portugal}

\author{Franz Gross }
\email[]{gross@jlab.org}
 \affiliation{ Thomas Jefferson National Accelerator Facility (JLab), 
Newport News, Virginia 23606, USA}

\date{\today}

\begin{abstract}
The $\pi$-$\pi$ scattering amplitude calculated with a model for the 
quark-antiquark interaction in the framework of the Covariant Spectator 
Theory (CST) is shown to satisfy the Adler zero constraint imposed by 
chiral symmetry. The CST formalism is established in Minkowski space and our calculations 
are performed in momentum space. We prove that the axial-vector Ward-Takahashi 
identity is satisfied by our model. Then we show that, similar to what happens 
within the Bethe-Salpeter formalism, application of the axial-vector Ward-Takahashi 
identity to the CST $\pi$-$\pi$ scattering amplitude allows us to sum the intermediate 
quark-quark interactions to all orders. The Adler self-consistency zero for $\pi$-$\pi$ 
scattering in the chiral 
limit emerges as the result for this sum. 

\end{abstract}


\maketitle

\section{\label{sec:intro} Introduction }

In the present scenario of both experimental and theoretical hadron physics 
the pion remains an important system to trace signatures of QCD in empirical observables.
The importance of the pion is multifaceted:  it emerges non-perturbatively as 
a quark-antiquark bound state, 
it is the Goldstone-boson mode associated with Spontaneous Chiral-Symmetry 
Breaking (S$\chi$SB), and it also contributes significantly, through the formation of a pion cloud, to the structure of the nucleon and to its coupling to external photons. In addition, the exchange of pions dominates the interaction between nucleons at larger distances and gives rise to a tensor force that strongly influences the structure of nuclei.

Traditionally, the non-perturbative dynamics
underlying hadronic systems have been addressed from two different perspectives, 
constituent quark models~\cite{Godfrey,Eichten:1975,Eichten:1978,Richardson:1978bt} and  QCD sum rules. These approaches, however, cannot provide 
a unified description of light mesons and baryons, nor can they avoid a delicate 
fine-tuning between a large number of parameters.  More recently, QCD simulations on 
the lattice~\cite{Edwards,Guo}, light-front formulations of quantum field theory~\cite{Brodsky:1997de,Carbonell:1998rj,Sales:1999ec}, 
as well as models based on the Dyson-Schwinger approach and mass gap equation~\cite{Bars:1977ud,Amer:1983qa,LeYaouanc:1983it,Bicudo:1989sh,Bicudo:1989si,Bicudo:1989sj,Nefediev:2004by,Alkofer:2000wg,Maris:2003vk,Fischer:2006ub,Rojas:2013tza}, have contributed to a more 
integrated perspective of mesons and baryons.

In particular, the Dyson-Schwinger framework generates dynamical quark models 
where the dressed quark mass is calculated as a function of the momentum, and moreover, this dynamical generation of quark masses is made consistent with the two-body quark-antiquark dynamics. However, lattice QCD and Dyson-Schwinger equations are usually solved in Euclidean space. In contrast, the
Covariant Spectator Theory (CST), used in this paper, works in Minkowski space, and also exhibits these features. 

First model 
calculations of the pion form factor using the solutions of the 
CST-Bethe Salpeter equation (CST-BSE) and the CST-Dyson equation (CST-DE) were presented in 
Ref.~\cite{pion_form_factor_paper}. There, the CST interaction kernel in momentum space
was taken as a $\delta$-function {\it plus} a covariant generalization of 
the linear confining interaction. 

The confining part in momentum space contains an important subtraction term that makes sure that it reduces to the linear potential (in coordinate space) $V_L(r)\propto r$ in the 
nonrelativistic limit. In particular,
it was seen in Ref.~\cite{Gross:1991pk} that the condition  $V_L(r=0)=0$ implies
that the confinement interaction decouples from 
the CST-DE for the scalar part of the dressed quark propagator, as well 
as from the CST-BSE for a massless pion in the chiral limit. 
For a scalar confining interaction, this decoupling property of our CST model is 
a necessary condition to ensure consistency with chiral symmetry. For the numerical 
predictions, our model 
was calibrated by adjusting the dressed quark mass function to the existing lattice 
QCD data. 

In this paper we submit our model to a more stringent test. We present the CST 
calculation of the $\pi$-$\pi$ scattering amplitude in the chiral limit, and conclude 
that it satisfies the Adler self-consistency zero as imposed by chiral symmetry, see Ref.~\cite{Adler_PhysRev.137.B1022}, 
 provided the  interaction kernel satisfies the Axial-Vector Ward-Takahashi Identity 
(AV-WTI). There are various possible choices for the Dirac structure of the kernel that satisfy
the AV-WTI. We choose a mixture of scalar, pseudoscalar, vector, axial-vector, and tensor
structures for the confining interaction, in  combination with a 
 vector--axial-vector structure for the remaining part of the kernel. Although one lacks first-principle evidence for 
scalar quark confinement, it is still quite important to study to what extent such 
confining forces can be made compliant with S$\chi$SB. To this effect the 
 AV-WTI will play a fundamental role when it comes to evaluate, 
to all orders of kernel insertions and independently of parameter fixing, $\pi$-$\pi$ scattering and the 
corresponding $\pi$-$\pi$ Adler zero.       

This paper is organized as follows: In Section \ref{sec:intro1} a brief 
review 
of the CST 
formalism is given. In Section \ref{sec:intro2} we discuss the constraints 
imposed by the AV-WTI on the CST interaction kernel and we specify the particular 
form of the kernel to be used in this paper. In Section \ref{sec:Adler} we present 
a calculation of $\pi$-$\pi$ scattering, first in the simple impulse approximation 
that is seen not to comply with the Adler zero in the chiral limit, and then to all 
orders in intermediate interactions that does yield the Adler-zero. Finally, 
in Section \ref{sec:summary} we present a brief summary and  our main conclusions.

\section{\label{sec:intro1} Brief review of the CST framework}
\,

The purpose of this section is to briefly review the basic ideas of the 
CST when applied to quark-antiquark mesons~\cite{Savkli:1999me,PhysRevD.89.016005}.
First, let us consider the four-dimensional BSE~\cite{Sal51} for 
heavy-light mesons. It is well known~\cite{Gross:1993zj} that cancellations 
occur between  iterations of ladder diagrams and higher-order crossed-ladder diagrams in the complete kernel of the BSE.  Owing to this, the omission of crossed-ladder diagrams and of certain 
pole contributions of the ladder diagrams from the kernel can actually yield a better approximation 
to the exact BSE than the ladder approximation does. 

This fundamental idea of CST 
emerges more formally from reorganizing the Bethe-Salpeter series, with a 
complete kernel and (off-mass-shell) 
two-particle propagators, into an equivalent form---the CST equation---where \textit{both} the kernel and propagators in the intermediate 
states are redefined. In the heavy-light case, the new quark propagators are 
chosen in such a way that, when the new kernel is truncated, only the positive-energy pole contribution from the heavy quark propagator 
in the energy loop integration is kept, which effectively corresponds to taking  
the heavy quark to be on its  positive-energy 
mass shell. 

The resulting three-dimensional equation, the one-channel CST (or Gross) 
equation~\cite{Gro69}, is manifestly covariant. But, unlike the BSE in ladder approximation, the CST equation also has a smooth nonrelativistic limit, and it can thus be viewed as a 
natural covariant extension of the quantum mechanical Dirac and 
Schr\"{o}dinger equations to quantum field theory. While the simple CST equation is very efficient for the description of 
heavy-light mesons, in the case of light quarks an explicitly 
charge-conjugation-symmetrized CST-BSE must be used. 
This is the case for the pion where the vertex 
functions of $\pi^+$ and $\pi^-$ are connected by charge conjugation and, therefore, both positive- and 
negative-energy quark poles must be included. 

 \begin{figure*}
\includegraphics[clip=0cm,width=9cm]{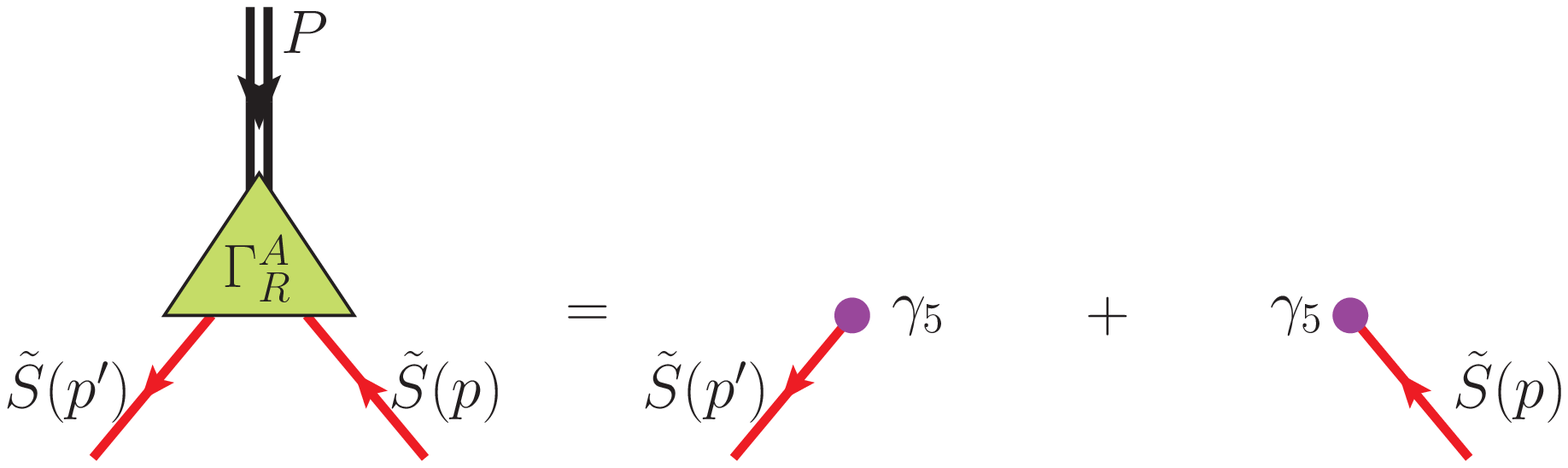}\\\vspace{0.5cm}
\includegraphics[clip=0cm,width=12cm]{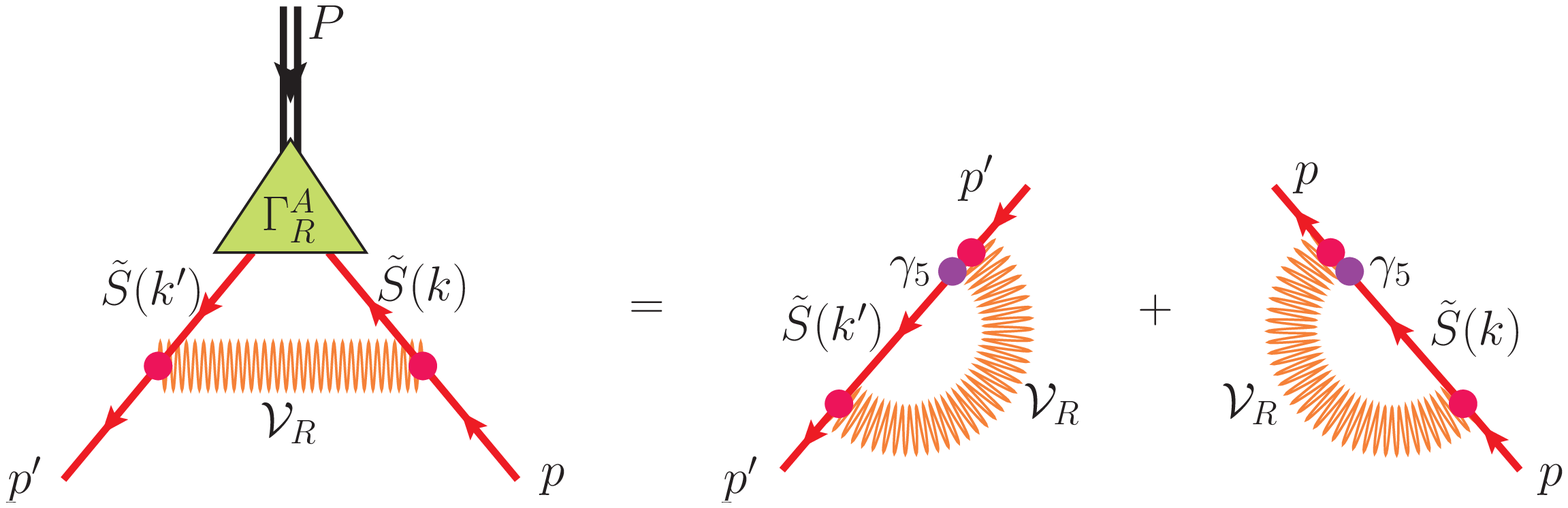}
\caption{(Color online) Top panel: The AV-WTI illustrated diagrammatically 
for the dressed current. Bottom panel: representation of the rhs of 
Eq. (\ref{eq:CSTBSEGammaA}) after application of the AV-WTI from the top panel. Each red arrowed line denotes a dressed quark propagator. The purple blobs denote $\gamma^5$ matrices and the pink blobs denote the Dirac 
structure of the kernel.
}\label{fig:AVWTI}
\end{figure*}

The idea of symmetrizing over 
all quark poles generates %
 the charge-conjugation-symmetric CST-BSE \cite{PhysRevD.89.016005},
 \linebreak
 \begin{widetext} 
\bea
\Gamma(p_1,p_2)=-&&\frac12Z_0 \int_k
\Big[{\cal V}(p,\hat k-\sfrac12  P)\Lambda(\hat k)\Gamma(\hat k,
\hat k-  P)S(\hat k-  P)
+{\cal V}(p,\hat k+\sfrac12  P) S(\hat k+  P)\Gamma(\hat k+  P,  
\hat k)\Lambda(\hat k)
\nonumber\\&&
+{\cal V}(p,-\hat k-\sfrac12  P) \Lambda(-\hat k)\Gamma(-\hat k,  -
\hat k-P)S(-\hat k-  P)
+{\cal V}(p,-\hat k+\sfrac12  P) S(-\hat k+  P)\Gamma(-\hat k+  P,  -
\hat k)\Lambda(-\hat k)\Big]\,
\nonumber\\\equiv&& \,\mathrm i\int_{k0}
{\cal V}(p,k) 
S(k+\sfrac P2)\,\Gamma(k+\sfrac P2,k-\sfrac P2)\,S(k-\sfrac P2)\,, 
\label{eq:CST-BSE}
\eea
\end{widetext}
where we use the short-hand notation for the three-dimensional covariant integration volume element,
\bea
\int_{\peq {k}} \equiv \int  \frac{\mathrm d^3 k}{(2\pi)^3} \frac{m}{E_k}\,,
\eea
and the last line of Eq.~(\ref{eq:CST-BSE}) introduces the notation \lq \lq $k0$'' 
to indicate the charge-conjugation invariant CST prescription for performing the $k_0$ contour integration. This amounts to keeping the average of the four propagator 
pole contributions from closing the contour in \textit{both}, the upper and the 
lower half-complex $k_0$ plane (for more details see Ref.~\cite{PhysRevD.89.016005}). 
With these definitions,
\bea
\mathrm i\int_{k0} \equiv \mathrm i\int\frac{\mathrm d^4k}{(2\pi)^4}\,\bigg|_{\footnotesize\begin{array}{l}k_0\;\text{propagator} 
\cr \text{poles only}\end{array}}=-\frac12\sum_{\footnotesize\begin{array}{c}\text{propagator}
\cr \text{pole terms}\end{array}}\int_{\peq k} \,.\nonumber\\\label{eq:pp}
\eea
The quantities in Eq.~(\ref{eq:CST-BSE}) are:  $\Gamma (p_1,p_2)$,  the ($4\times 4$) bound-state 
vertex function
with $p_1=p+\sfrac P2$  and $-p_2=-p+\sfrac P2$ the four-momenta of the outgoing quark and antiquark (respectively);   $P$,  the total bound-state momentum;  
 $\hat k=(E_k,{\bf k})$,  the on-shell 
four-momentum with $E_k=\sqrt{m^2+{\bf k}^2}$;
 $\mathcal V (p,k)\equiv \mathcal V (p,k;P)$, the interaction 
kernel; $S(k)$, the dressed quark  propagator; and $\Lambda(k)=[M(k^2)+\slashed k]/2 M(k^2)$ where $M(k^2)$ is the dressed quark mass function. The kernel is an operator, and we use the shorthand notation 
 \bea
 {\cal V}(p,k) {\cal X}\equiv\sum_i V_i(p,k){\cal O}_i{\cal X} {\cal O}_i \,,\label{eq:Vdecomp}
 \eea
 where the sum $i=\{ S, P, V, A, T\}$ is over the five possible invariant structures that could contribute: scalar, pseudoscalar, vector, axial-vector, and tensor.  This will be discussed further when it is needed below.  
The dressed quark propagator is given by    
\bea
S(p)=\frac1{m_{0}-\slashed{p}+\Sigma(p)-\mathrm i\epsilon}\, , 
\label{eq:dressedprop}
\eea
where $m_{0}$ is the bare quark mass and $\Sigma (p)$ is the quark self energy, which is the 
 solution of the one-body CST-DE involving, 
for consistency, the \textit{same} interaction kernel $\mathcal V$ that 
dresses the quark-antiquark vertex. The CST-DE is 
given by~\cite{PhysRevD.89.016005} 
\bea
\Sigma (p)&=& \frac12Z_0\int_k
\Big\{{\cal V}(p,\hat k)\Lambda(\hat k)+{\cal V}(p,-\hat k)\Lambda(-\hat k)\Big\}
\nonumber\\&\equiv& -\mathrm i\int_{k0}
{\cal V}(p,k) S(k)\,.\label{eq:CST-DE}
\eea
Writing the self-energy in the form
\bea
\Sigma(p)=A(p^2)+\slashed{p} B(p^2)
\eea
leads to a dressed propagator of the form
\bea\label{eq:dressedprop1}
S(p)= Z(p^2)\frac{M(p^2)+\slashed{p}}{M^2 (p^2)-p^2-\mathrm i\epsilon}\,,
\eea
where the mass function $M(p^2)$ and the wave function normalization $Z(p^2)$ are
\bea
M(p^2)&=&\frac{A(p^2)+m_0}{1-B(p^2)}\,,
\nonumber\\
Z(p^2)&=&\frac{1}{1-B(p^2)}\,,
\eea
and $Z_0\equiv Z(m^2)$. For $\Sigma (p)=0$, $S(p)$ becomes the bare propagator denoted as $S_0(p)$. 

A proof of principle that the CST-Bethe-Salpeter Eq.~(\ref{eq:CST-BSE})  and the CST-Dyson Eq.~(\ref{eq:CST-DE}) are actually numerically 
manageable  in Minkowski space and that they underlie a dynamical 
quark model that incorporates 
S$\chi$SB (similar to the Dyson-Schwinger approach)  was presented 
in Refs.~\cite{pion_form_factor_paper,PhysRevD.89.016005}. In this paper we build on the model introduced 
in those recent references, where technical details can be found.  

\begin{figure*}
\includegraphics[clip=0cm,width=12cm]{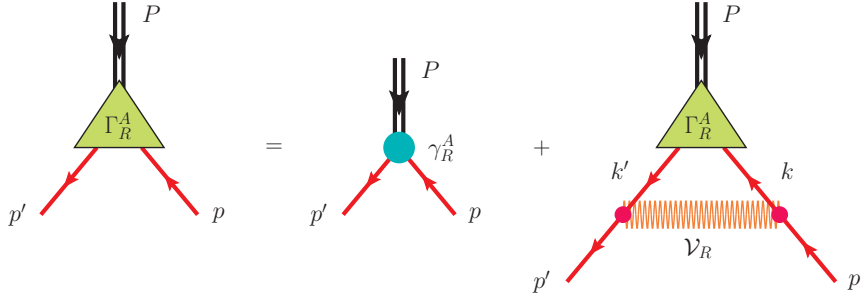}
\caption{(Color online) The inhomogeneous CST-BSE for 
$\Gamma^{A}_R (p^\prime,p)$.
}\label{fig:BSEaxial}
\end{figure*}

It was already proven in Refs.~\cite{Gross:1991te,PhysRevD.89.016005} that in the chiral 
limit the pion mass also vanishes, which means that the CST equations are, at least 
at this level, not inconsistent with the requirements of dynamical chiral symmetry 
breaking. In the present work we look at the implications of chiral symmetry on 
our model coming from the AV-WTI and $\pi$-$\pi$-scattering in the chiral limit.

\section{\label{sec:intro2} The axial-vector Ward-Takahashi identity and the interaction 
kernel}
\subsection{The axial-vector Ward-Takahashi identity}
In  our previous work, Ref.~\cite{PhysRevD.89.016005}, the quark-antiquark 
interaction was regularized by a strong quark form factor $h$ associated with each quark line entering or leaving a vertex. These form factors can 
be moved from the interaction vertices to the quark propagators, which leads to the replacement of the original kernel ${\mathcal V}(p,k)$ by a  \textit{reduced} kernel ${\mathcal V}_R(p,k)$, dressed propagators $S(p)$ by \textit{damped} dressed propagators $\tilde S(p)=h^2 (p^2) S(p)$, and bare propagators $S_0(p)$ by \textit{damped} bare propagators $\tilde S_0(p)=h^2 (p^2) S_0(p)$.
We use reduced kernels that depend only on the square of the transferred momentum, such that
\begin{eqnarray}
 \mathcal V_{R}(p-k)=h^{-1}(p^2) h^{-1}(p'^2) h^{-1}(k^2) h^{-1}(k'^2)
\mathcal V(p,k)\,,\nonumber\\
\end{eqnarray}
 where $\mathcal V(p,k)$ is the kernel of Eq.~(\ref{eq:CST-BSE}).

Chiral symmetry and its breaking is expressed through the AV-WTI, which can be derived from the divergence of the axial-vector current~\cite{Adler:1969gk}. Expressed in terms of the reduced vertex functions and the damped propagators, the familiar AV-WTI for off-shell quarks is
\bea
P_\mu \Gamma^{5\mu}_R(p',p)+2m_0 \Gamma^5_R(p',p)&=&\tilde S^{-1} (p')\gamma^5+\gamma^5\tilde S^{-1} (p)
\nonumber\\
&\equiv&\Gamma^{A}_R (p^\prime,p) \,,\label{eq:AVWTI}
\eea 
where $\Gamma^{5\mu}_R(p',p)$ is the reduced dressed axial-vector vertex, $\Gamma^5_R(p',p)$ the reduced dressed pseudoscalar vertex, $p$ and $p'$ are the incoming and outgoing quark momenta, respectively, and $P=p'-p$ is the momentum flowing into the vertex to which the incoming and outgoing quarks connect. The quantity $\Gamma^{A}_R (p^\prime,p)$ defined by the lhs of Eq.~(\ref{eq:AVWTI}), which we refer to as the \lq\lq axial vertex'', is a convenient combination of the axial-vector and the pseudoscalar vertices used in Refs.~\cite{PhysRevC.67.035201,PhysRevD.65.076008}. The identity~(\ref{eq:AVWTI}) is illustrated in the upper panel of Figure~\ref{fig:AVWTI}. Note that the AV-WTI for bare quark propagators $S_0$ implies that the bare axial-vector and pseudoscalar vertices are $\gamma^5\gamma^\mu$ and $\gamma^5$, respectively.

The dressed axial-vector vertex, the dressed pseudoscalar vertex, and the dressed axial vertex are all solutions of an inhomogeneous CST-BSE.  For the axial vertex, 
\begin{eqnarray} 
\Gamma^{A}_R (p^\prime,p)&=&\gamma^{A}_R (p^\prime,p)\nonumber\\&&+\mathrm i
\int_{k0} \mathcal V_R(p-k) 
\tilde S(k') \Gamma^{A}_R (k',k) \tilde S(k) \,, \qquad  \label{eq:CSTBSEGammaA}
\end{eqnarray}
where $\gamma^{A}_R (p^\prime,p)$ is the reduced bare axial vertex (to be discussed below) and  Eq.~(\ref{eq:CSTBSEGammaA}) is depicted diagrammatically 
in Fig.~\ref{fig:BSEaxial}. Note that $p-k=p'-k'$. The dressed damped propagator $\tilde S(p)$ is the solution of the CST-DE 
\begin{eqnarray}
 \tilde S^{-1} (p)=\tilde S_0^{-1} (p)-\mathrm i\int_{k0} 
{\cal V}_R(p-k) \tilde S(k)\, .
\label{eq:CST-DE2}
\end{eqnarray}

\begin{widetext}
Next, look at the implications of the AV-WTI (\ref{eq:AVWTI}) and how it relates to the one-body CST-DE (\ref{eq:CST-DE2}) and to the inhomogeneous two-body
CST-BSE (\ref{eq:CSTBSEGammaA}). Using Eq.~(\ref{eq:AVWTI}) in the integrand of Eq.~(\ref{eq:CSTBSEGammaA}),  the result splits into two terms
\begin{eqnarray}
\Gamma^{A}_R (p^\prime,p)&=&\gamma^{A}_R (p^\prime,p)+\nonumber\mathrm i
\int_{k0}\mathcal V_R(p-k) \tilde S(k') \left[\tilde S^{-1} (k')\gamma^5+
\gamma^5\tilde S^{-1} (k)\right] \tilde S(k)\nonumber\\&=&\gamma^{A}_R (p^\prime,p)+
\mathrm i\int_{k0}
\mathcal V_R(p-k)  \gamma^5 \tilde S(k)+ \mathrm i\int_{k0}\mathcal V_R(p'-k')\tilde S(k')\gamma^5 \,,
\label{eq:AVWTI2} 
\end{eqnarray}
where $p'=P+p$ and $k'=P+k$, as illustrated in the lower panel of Fig.~\ref{fig:AVWTI}. On the other hand, with the shorthand notation $h=h(p^2)$ and $h'=h(p'^2)$, the rhs of Eq.~(\ref{eq:AVWTI}) reads
\begin{eqnarray}
\frac{m_0-\slashed p'+\Sigma (p')}{h'^2}
\gamma_5+\gamma_5\frac{m_0-\slashed p+\Sigma (p)}{h^2}
&=& \gamma^5\left[\frac{\slashed p'}{h'^2}-\frac{\slashed p}{h^2}\right]
 +m_0\left[\frac{1}{h'^2}+\frac{1}{h^2}\right]\gamma_5+
\frac{\Sigma (p')}{h'^2}\gamma_5+\gamma_5 
\frac{\Sigma ( p)}{h^2}
 \, ,
 \label{eq:AVWTI3}
\end{eqnarray}
\end{widetext}
and we can now compare  Eq.~(\ref{eq:AVWTI2}) with (\ref{eq:AVWTI3}).

\subsection{The reduced bare vertex}\,
First, on both sides we identify  all quantities that do not involve 
contributions to the self-energy from the dynamical dressing by the kernel, 
and conclude that
\begin{eqnarray}
\gamma^{A}_R (p^\prime,p)&=&\gamma_5\Big[\frac{\slashed p'}{h'^2}-\frac{\slashed p}{h^2}\Big]
 +m_0\left[\frac{1}{h'^2} +\frac{1}{h^2}\right]\gamma_5\nonumber\\
& = &  \tilde S_0^{-1} (p')\gamma^5+\gamma^5
\tilde S_0^{-1} (p) \,.\qquad\label{eq:baredampedAVWTI}
\end{eqnarray}
This is an AV-WTI for the damped bare vertex and propagators.  It can be 
satisfied by decomposing $\gamma^A_ R(p',p)$  into its pseudoscalar and 
axial-vector parts, 
\begin{eqnarray}
\gamma^A_ R(p',p)= P_\mu \gamma^{5\mu}_R(p',p)+2m_0 
\gamma^{5}_R(p',p) \, ,
\label{eq:gammaA}
\end{eqnarray}
and making the following simple \textit{Ans\"{a}tze} for $\gamma^{\mu5}_R$ and $\gamma^{5}_R$ in the manner of Refs.~\cite{Gro87,Gro93,Gro96}:
\begin{eqnarray}
 \gamma^{5\mu}_R (p^\prime,p)&=&f_A(p^\prime,p)\gamma^5\gamma^\mu
 \nonumber\\&&+ 
g_A(p^\prime,p) \Lambda_0(-p') \gamma^5\gamma^\mu
\Lambda_0(-p)\label{eq:gammaAR5mu}
\end{eqnarray}
and
\begin{eqnarray}
 \gamma^{5}_R (p^\prime,p)&=&f_P(p^\prime,p)\gamma^5
 \nonumber\\
 &&+g_P(p^\prime,p) 
\Lambda_0(-p') \gamma^5\Lambda_0(-p)\,,\label{eq:gammaAR5}
\end{eqnarray} 
where  $\Lambda_0(p)=(m_0+\slashed p)/2m_0$. The form factors
$f_A(p^\prime,p)$, $g_A(p^\prime,p)$, $f_P(p^\prime,p)$, and $g_P(p^\prime,p)$ are then
determined to be 
\begin{eqnarray}\label{eq: 1}
 f_A(p',p)&=&f_P(p',p)=\frac{m_0^2- p'^2}{ h'^2 (p^2- p'^2)}-
\frac{ m_0^2-p^2}{h^2 (p^2- p'^2)}\qquad
\\ \label{eq: 2}
 g_A(p',p)&=&-g_P(p',p)=\frac{4 m_0^2 }{p'^2-p^2}\left(\frac{1}{h^2}-
\frac{1}{h'^2}\right)\,.
\end{eqnarray}
If we set all quark 
form factors $h$  to 1, then $f_A(p',p)\rightarrow1$ and $g_A(p',p)\rightarrow0$, such that $\gamma^{5\mu}_R(p',p)
\rightarrow \gamma^5\gamma^\mu$ and $\gamma^{5}_R(p',p)\rightarrow \gamma^5$. In this case, 
the damped bare AV-WTI~(\ref{eq:baredampedAVWTI}) becomes the bare one involving $S_0$, as used, for instance, in Ref.~\cite{PhysRevD.65.076008}.

\subsection{Constraints on the interaction kernel}\label{Sec:kernel}
 
The aim of this subsection is to determine the general form of the covariant 
interaction kernel $\mathcal V_R(p-k)$  such that  
the AV-WTI~(\ref{eq:AVWTI}) is  satisfied. Recalling the decomposition (\ref{eq:Vdecomp}), the reduced kernel will be written in the form 
\bea
{\cal V}_R(p-k)&=& V_{SR}(p-k){\bf 1}\otimes{\bf 1}+V_{PR}(p-k)
\gamma^{5}\otimes \gamma^{5}\nonumber\\&&\,\,+V_{VR}(p-k)
\gamma^\mu
\otimes \gamma_{\mu}\nonumber\\&&\,\,+ V_{AR}(p-k)
\gamma^5
\gamma^\mu\otimes \gamma^5\gamma_\mu\nonumber\\&&\,\,+ \frac12V_{TR}(p-k)
\sigma^{\mu\nu}\otimes \sigma_{\mu\nu}
\,,
\label{eq:BSkernel}
\eea
where the corresponding factors in the decomposition of $\mathcal V(p,k;P)$ include 
the strong quark form factors and are therefore
$V_i(p,p';k,k')=h(p^2) h(p'^2)h(k^2) h(k'^2) V_{iR}(p-k)$.  Using this decomposition, comparing Eqs.~(\ref{eq:AVWTI2}) 
and~(\ref{eq:AVWTI3}), and extracting the $\gamma^5$, we see that preserving the AV-WTI is tantamount to requiring that
\be\label{eq:dynAVWTI1}
 \mathrm i\int_{k0}
\widehat{\mathcal  V}_R(p-k)  \tilde S(k)= 
\frac{\Sigma(p)}{h^2(p^2)},
\ee 
where the operator  $\widehat{\mathcal  V}_R(p-k)$ is obtained from the operator ${\mathcal  V}_R(p-k)$ by changing the sign of the vector and axial-vector components of ${\mathcal  V}_R(p-k)$. Using (\ref{eq:BSkernel}), (\ref{eq:dynAVWTI1}) reduces to
\begin{eqnarray}
\frac{\Sigma (p)}{h^2(p^2)}=  \mathrm i\,  
\int_{k0} &&\Big[ V_{SR}(p-k) 
\tilde S(k)+ V_{PR}(p-k) 
\gamma^5\tilde S(k)\gamma^5 \nonumber\\&&- V_{VR}(p-k) \gamma^\mu 
\tilde S(k) \gamma_{\mu}\nonumber\\&&- V_{AR}(p-k) \gamma^5\gamma^\mu 
\tilde S(k) \gamma^5\gamma_{\mu}\nonumber\\&&+\frac12 V_{TR}(p-k) 
\sigma^{\mu\nu}\tilde S(k)\sigma_{\mu\nu}\Big]
\,.\label{eq:Sigma}
\end{eqnarray}
By comparing this equation with the one-body CST-DE for the self-energy, 
Eq.~(\ref{eq:CST-DE}), using the same kernel we conclude, given the signs in front of the scalar, pseudoscalar, and tensor 
interaction terms in (\ref{eq:Sigma}), that the AV-WTI links the one-body 
CST-DE with the two-body CST-BSE, Eq.~(\ref{eq:AVWTI2}), if and only if
\begin{eqnarray}
 \int_{k0} &&\Big[V_{SR}(p-k) 
\tilde S(k)+ V_{PR}(p-k)\tilde S(-k)\nonumber\\&&\,\,+\frac12 V_{TR}(p-k) 
\sigma^{\mu\nu} \tilde S(k)\sigma_{\mu\nu}\Big]=0\,.\label{eq:condV_SP}
\end{eqnarray}
In the literature, the most common realization of this type of condition is achieved 
by setting $V_{SR}(p-k)=V_{PR}(p-k)=V_{TR}(p-k)=0$, i.e. by using only interaction kernels that anticommute with $\gamma^5$, like vector or axial-vector (e.g., see Refs.~\cite{Adler:1984ri,PhysRevC.56.3369,PhysRevD.65.076008}). 

In this
work we use a kernel that \textit{does} include nonvanishing scalar, 
pseudoscalar, and tensor structures, but is, nevertheless, consistent with the AV-WTI. Other models with this feature exist in the literature. In Ref.~\cite{PhysRevD.47.1145}, a tensor term was chosen in such a way that Eq.~(\ref{eq:condV_SP}) is satisfied. 
In our case, it is the implementation of linear confinement in the CST framework that makes sure Eq.~(\ref{eq:condV_SP}) holds, with or
without a tensor term. 

\subsection{Linear confinement}\label{sec:confinement}
In this section we specify the momentum-dependent parts of the kernel and we 
discuss, in particular, how confinement is implemented in our CST model.
In the literature there are several examples of confinement potentials. For instance,
it is well-known that
the static potential in the quenched approximation of lattice QCD can be 
parameterized by a Cornell-type potential~\cite{Allton:2001sk}.

However the chiral limit is quite different from the quenched limit. In this paper we want to investigate how a linear scalar confinement can be made
compatible with S$\chi$SB. Other Dirac structures are possible, namely vector confinement, but here we choose scalar confinement as the most stringent case still able to hold
the phenomenology of chiral symmetry.

To this end, we implement linear confinement in a relativistically 
generalized form of the momentum-dependent kernel functions $V_{iR}$ in Eq.~(\ref{eq:BSkernel}). The confinement part of the $V_{iR}$'s is denoted  $V_L$, and its action on an arbitrary function $\phi$ of the off-shell quark momentum $p$, in the one-body CST-DE, is given by
\bea
\langle V_L\phi\rangle(p)&=&\frac12\int_{k} V_{A}(p,\hat k) \left[\phi(\hat k)-
\phi(\hat p_R)\right]\nonumber\\&&+ \frac12\int_{k}V_{A}(p,-\hat k) \left[\phi(-\hat k)-
\phi(\hat p_R)\right]\, ,
\label{eq:VL}
\label{eq:VLrelativistic}
\eea
where \bea 
V_{A}(p,\hat k)= -h^2(p^2)h^2(m^2)
\frac {8\pi\sigma}{(p-\hat k)^4} \, ,
\label{eq:VA}
\eea
and $\hat k$ is the on-shell quark momentum in the loop integral. 
The subtraction term, $\phi(\hat p_R)$, regularizes the singularities of $V_{A}$ at $(\hat k-\hat p)^2=0$. 
The argument of the subtraction term is $\hat p_R=(E_{p_R},{\bf p}_R)$, 
where ${\bf p}_R={\bf p}_R(p_0,\bf p)$ are the values of $\bf k$ at which 
either $V_{A}(p,\hat k)$ or  $V_{A}(p,-\hat k)$ become singular. 

When applied to the wave function $\Psi(p_1,p_2)$ of a two-quark system depending on the two quark momenta $p_1=p+P/2$ and $p_2=p-P/2$, the action of $V_L$  is defined by
\bea
&&\langle V_L\Psi\rangle(p_1,p_2)\nonumber\\&&=\frac 12\int _{ k}\Big\{
V_{A}(p,\hat k-\sfrac P2) \left[\Psi(\hat k,\hat k-P)-
\Psi(\hat p_{R1}^+,\hat p_{R1}^+-P)\right]\nonumber\\&&\,+ 
 V_{A}(p,\hat k+\sfrac P2) \left[\Psi(\hat k+P,\hat k)-
\Psi(\hat p_{R2}^++P,\hat p_{R2}^+)\right]\nonumber\\&&\,+
 V_{A}(p,-\hat k-\sfrac P2) \left[\Psi(-\hat k,-\hat k-P)-
\Psi(-\hat p_{R1}^-,-\hat p_{R1}^--P)\right]\nonumber\\&&\,+ 
 V_{A}(p,-\hat k+\sfrac P2) \left[\Psi(-\hat k+P,-\hat k)-
\Psi(-\hat p_{R2}^-+P,-\hat p_{R2}^-)
\right]\Big\}\,,\nonumber\\
\label{eq:VLrelativistic2}
\eea
where now
\bea 
V_{A}(p,k)= -h(p_1^2)h(p_2^2)h(k_1^2)h(k_2^2)
\frac {8\pi\sigma}{(p-k)^4} \,, 
\label{eq:VA2}
\eea
with $k_1=k+P/2$, $k_2=k-P/2$, and ${\bf p}_{R1}^\pm$ and ${\bf p}_{R2}^\pm$ being the values of ${\bf k}$ at which $V_{A}(p,\pm \hat k-\sfrac P2)$ and 
$V_{A}(p,\pm \hat k+\sfrac P2)$ become singular, respectively. The CST wave functions where one quark is on-shell 
are 
\bea 
\Psi(\hat p_1,p_2)&=&\Lambda (\hat p_1) \Gamma (\hat p_1,p_2) S(p_2)\,,\\
\Psi(p_1,\hat p_2)&=&S( p_1) \Gamma (p_1,\hat p_2) \Lambda(\hat p_2)\,.
\eea
The subtraction terms regularize both the \textit{diagonal} singularities of 
$V_{A}$ at $(\hat k-\hat p)^2=0$, i.e., in channels where the same quark is on mass-shell  in the initial and intermediate states, 
and the \textit{off-diagonal} singularities  
at $(\pm \hat k+P- \hat p)^2=0$, which occur in channels with different quarks on mass-shell in 
the initial and intermediate states.
The subtraction also leads directly to the important relation
 \begin{eqnarray}
\langle V_L\rangle(p)= \int_{k}  V_L(p,\hat k)=0 \, ,
\label{eq:VLzero}
\end{eqnarray} 
which is a relativistic generalization of the nonrelativistic $V_L (r=0)=0$.

Equation (\ref{eq:VLzero}) allows the use of scalar, pseudoscalar, and tensor confining interactions  in a way that is still consistent with chiral symmetry, because it makes it possible to 
satisfy Eq.~(\ref{eq:condV_SP}). How this works in detail 
will be addressed shortly.

As a consequence of Eq.~(\ref{eq:VLzero}), the linear confinement $V_{L}$ does not contribute to the scalar 
part of the self-energy $A(p^2)$ [which means that $A_L(p^2)=0$], nor 
to the pion equation in the chiral limit \cite{Gross:1991pk}. Therefore, a 
scalar component in the confinement potential is not necessarily inconsistent 
with chiral symmetry.

\begin{figure*}
\includegraphics[clip=0cm,width=9cm]{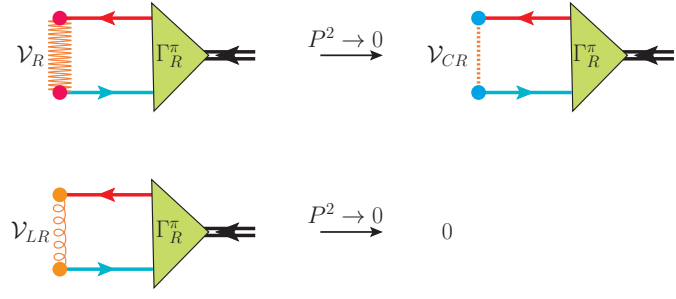}
\caption{(Color online) In the chiral limit of vanishing pion mass only the 
non-confining part of the kernel contributes to the pion CST-equation. Each red or blue arrowed line denotes a dressed quark propagator. The light-blue and dark-yellow blobs denote the Dirac structures of $\mathcal V_{CR}$ and $\mathcal V_{LR}$, respectively. 
}\label{fig:pionChL}
\end{figure*}

To discuss the implications of the AV-WTI on the kernel let us specify the Dirac 
structure of $\cal V_R$  as follows:  
\begin{eqnarray}\label{eq:vectorpart}
 && V_{SR}(p-k)=\lambda_S V_{LR}(p- k), \nonumber\\
 &&V_{PR}(p-k)=\lambda_P V_{LR}(p- k),\nonumber\\
 &&V_{VR}(p-k)=\lambda_V V_{LR}(p-k)+ \kappa_V V_{CR}(p- k),\nonumber\\
&&V_{AR}(p-k)=\lambda_A V_{LR}(p-k)+ \kappa_A V_{CR}(p- k),\nonumber\\
&&V_{TR}(p-k)=\lambda_T V_{LR}(p-k)
\,.
\end{eqnarray}
Here $V_{LR}$ is the reduced version of $V_L$, and $V_{CR}$ is a Lorentz invariant function representing the non-confining part of the interquark 
interaction of Eq.~(\ref{eq:BSkernel}).  The weight parameters $\lambda_i$ and $\kappa_i$ are constants. For a pure vector--axial-vector kernel, with $\lambda_S=\lambda_P=\lambda_T=0$, Eq.~(\ref{eq:condV_SP}) is trivially satisfied. However, a non-trivial realization is also possible. To obtain this, insert $V_{SR}$, $V_{PR}$, and $V_{TR}$ of 
(\ref{eq:vectorpart}) into~(\ref{eq:condV_SP}) and separate scalar and vector parts:
\begin{eqnarray}
  &&\int_{k0} V_{SR}(p-k) 
\tilde S(k)+ \int_{k0} V_{PR}(p-k) 
\tilde S (-k)\nonumber\\&&\,+\frac12\int_{k0} V_{TR}(p-k) 
\sigma^{\mu\nu} \tilde S(k)\sigma_{\mu\nu}
\nonumber\\&&\propto
 ( \lambda_S+\lambda_P+6\lambda_T)\int_{k} \left[ V_{LR}(p-\hat k)+ V_{LR}(p+\hat k)\right]  
\nonumber\\&&\,\,+
(\lambda_S-\lambda_P)\int_{k} \frac{\hat{\slashed k}}{m}
\left[V_{LR}(p-\hat k)-V_{LR}(p+\hat k)\right]\,.\label{eq:lslp}
\end{eqnarray}
According to Eq. (\ref{eq:VLzero}), the 
first integral vanishes because $\int_{k}  V_L(p,\hat k)=\int_{k}  V_L(p,- \hat k)=0$. For the second term to be zero we have to choose $\lambda_S=\lambda_P$, 
since the integral does not vanish. Note that $\lambda_T$ is not constrained by Eq.~(\ref{eq:lslp}) because the tensor part of the kernel does not contribute to the vector part of the self-energy. We conclude that a kernel that includes 
scalar linear confinement also requires an equal-weighted pseudoscalar counterpart, 
in order to satisfy the AV-WTI. Equation~(\ref{eq:VLzero}) implies that, in the chiral 
limit, only the non-confining part of the kernel, 
$\mathcal V_{C}(p,k)=[\kappa_V(\gamma^\mu
\otimes \gamma_{\mu})+\kappa_A(\gamma^5\gamma^\mu
\otimes \gamma^5\gamma_{\mu}) ] V_{C}(p,k)$, contributes to the massless pion equation. This is 
diagrammatically depicted in Fig.~\ref{fig:pionChL} and was proven in 
Ref.~\cite{PhysRevD.89.016005}. 

With the AV-WTI-preserving CST choice $\lambda_S=\lambda_P$ the contributions 
of the scalar and the pseudoscalar parts of the linear-confining kernel to the 
self-energy cancel exactly. \textit{Therefore, only the vector and 
axial-vector parts of the linear kernel contribute to the self-energy $\Sigma$, 
here denoted  $\Sigma_L$}. As one moves away from the chiral limit, the scalar, pseudoscalar, and tensor terms in the potential start to play a role in the pion equation.

Now, inserting Eqs~(\ref{eq:vectorpart}) into Eq.~(\ref{eq:CST-DE}) gives 
\vspace{0.1in}

\begin{widetext}
\begin{eqnarray}
&&\Sigma_L (p)=- \mathrm{i}\int_{k0} V_{VR}(p-k) 
\gamma^\mu \tilde S(k) \gamma_{\mu}-\mathrm{i}\int_{k0}  V_{AR}(p-k) \gamma^5\gamma^\mu 
\tilde S(k) \gamma^5\gamma_{\mu} \nonumber\\&&
 \propto 4 (\lambda_V-\lambda_A) \int_{k}  
\left[V_{LR}(p-\hat k) + V_{LR}(p+\hat k)\right]
-2(\lambda_V+\lambda_A) \int_{k} \frac{\slashed {\hat k}}{m} \left[V_{LR}(p-\hat k) -
V_{LR}(p+\hat k)\right] \,.\label{eq:lvla}
\end{eqnarray}
\end{widetext}
As in Eq.~(\ref{eq:lslp}), the first integral vanishes because $\int_{k}  V_L(p,
\hat k)=\int_{k} V_L(p,-\hat k)=0$. The second integral does not vanish and 
 contributes to the self-energy, unless $\lambda_V=-\lambda_A$. 

After this discussion of the general form of the interaction kernel, in the remainder of this paper we specialize to the particular case
\begin{eqnarray}
 \mathcal V_R(p-k)&=& V_{LR}(p-k)\Big[\lambda_S ({\bf 1}\otimes {\bf 1})+\lambda_S (\gamma^5\otimes\gamma^5) \nonumber\\&&+\lambda_V
( \gamma^\mu 
\otimes \gamma_{\mu})+ \lambda_A (\gamma^5\gamma^{\mu} \otimes \gamma^5\gamma_{\mu})\nonumber\\&&+
\frac{\lambda_T}{2}(\sigma^{\mu\nu}\otimes\sigma_{\mu\nu})\Big]  +V_{CR}(p-k)
\nonumber\\&& \times
\Big[\kappa_V(\gamma^\mu\otimes\gamma_\mu)+\kappa_A (\gamma^5\gamma^{\mu} \otimes \gamma^5\gamma_{\mu})\Big]\, .
\nonumber\\
\,\label{eq:kernel} 
\end{eqnarray}

\subsection{The pion vertex function and the axial vertex in the 
chiral limit}\,
Before we turn to $\pi$-$\pi$-scattering, it is useful to consider 
the implications of the AV-WTI on the pion and on the axial vertex functions in the chiral limit. 

\vspace{0.4in}
 
\subsubsection{Bare axial vertex}\,
We start with the reduced bare axial vertex $\gamma_R^A (p',p)$ as parametrized in Eqs.~(\ref{eq:gammaA})-(\ref{eq: 2}). In the chiral limit of vanishing bare quark mass, 
$m_0\rightarrow0$, and vanishing 
vertex momentum, $P^\mu\rightarrow0$, $f_A (p',p)$ remains finite whereas $g_A (p',p)$ 
vanishes, and thus the axial-vector vertex contracted with $P^\mu$ vanishes. 
For the remaining pseudoscalar part we have  for the form factors in the 
limit $P^\mu\rightarrow0$
\begin{eqnarray}
 f_P(p,p)=\frac{1}{h^2(p^2)}+\frac{2(m_0^2-p^2) }{h^3(p^2)} 
\frac{\mathrm d h(p^2) }{\mathrm d p^2} 
\end{eqnarray}
and 
\begin{eqnarray}
 g_P(p,p)=-\frac{8m_0^2}{h^3(p^2)} \frac{\mathrm d h(p^2) }{\mathrm d p^2}\,. 
\end{eqnarray}
The derivative 
terms  in $\gamma^{A}_R (p^\prime,p)$ of Eq.~(\ref{eq:gammaA}) cancel, and $\gamma^{A}_R (p,p)$ becomes
\begin{eqnarray}
 \gamma^{A}_R (p,p)=\frac{2m_0}{h^2(p^2)}\gamma^5\,, \label{eq:gammaAP0}
\end{eqnarray}
as it should according to Eq.~(\ref{eq:baredampedAVWTI}), and thus in the chiral limit
\begin{eqnarray}
 \lim_{\footnotesize\begin{array}{c}m_0\rightarrow0\cr p'\rightarrow p\end{array}} \gamma^{A}_R (p^\prime,p)=0\,.
\label{eq:gammaAchirall}
\end{eqnarray}

\begin{figure*}
\includegraphics[clip=0cm,width=11cm]{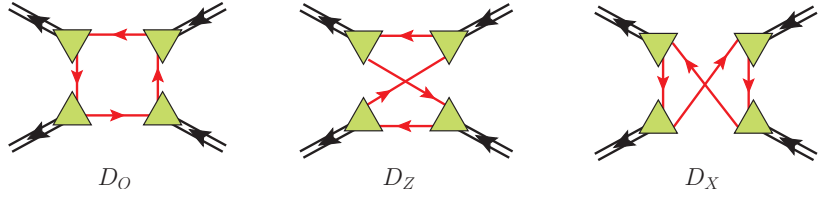}
\caption{(Color online) The direct contributions to $\pi$-$\pi$ scattering.  
}\label{fig:box_pipiscattering}
\end{figure*}

\subsubsection{Dressed axial vertex}
\label{pionvertex}
Because of Eq.~(\ref{eq:gammaAchirall}), the CST-BS equation~(\ref{eq:CSTBSEGammaA}) 
for $\Gamma^{A}_R$ 
becomes homogeneous in the chiral limit, and using the AV-WTI in the form of Eq.~(\ref{eq:AVWTI}), this  vertex function 
can be expressed directly in terms of the scalar mass function, $A(p^2)$
 \begin{eqnarray}
\Gamma^{A}_{R\chi} (p,p) 
=\gamma^5 
\frac{2A_\chi(p^2)}{h^2(p^2)}\,,\label{eq:GammaAchi}
\end{eqnarray}
where $A_\chi$ is the chiral limit of $A$.  Since a finite quark mass is generated by S$\chi$SB,  $A_\chi$ is non-zero, and it is clear 
from Eq.~(\ref{eq:GammaAchi}) that  
$\Gamma_{R\chi}^A (p,p)$ must also be finite in this limit. Note that the pion produces poles in both $\Gamma^{5}_R$ and 
$\Gamma^{5\mu}_R$, with the corresponding residues constrained to cancel through
the AV-WTI (for details, see, for instance Ref.~\cite{PhysRevD.65.076008}). In Ref.~\cite{PhysRevD.89.016005}, we found that the CST-BSE (\ref{eq:CST-BSE}) for a massless pion becomes identical to the scalar part of the CST-DE (\ref{eq:CST-DE}) in the chiral limit, provided the interaction kernel satisfies condition~(\ref{eq:condV_SP}). This implies, in particular, the relation
\begin{eqnarray}
 \Gamma^\pi_{R\chi}(p,p)&=&\gamma^5G_0Z_0\frac{A_\chi(p^2)}{m_\chi h^2(p^2)}\nonumber\\&=&
\frac{G_0 Z_0}{2m_\chi}\Gamma^{A}_{R\chi}(p,p)\,.
\label{eq:Gammachi}
\end{eqnarray}
Here the constant $G_0$ is the inverse norm of the pion vertex function $\Gamma^\pi_{R\chi}$, calculated from the triangle diagram for the pion form factor at zero-momentum transfer and $m_\chi$ is the dressed quark mass, obtained by solving the  equation $M_\chi(p^2=m_\chi^2)=m_\chi$ with $m_0=0$ and the strong quark form factors normalized to $h(m_\chi^2)=1$.

The next task is to use the AV-WTI to evaluate the $\pi$-$\pi$ scattering 
amplitude at threshold in the chiral limit, with the kernel iterated to all orders, 
and to obtain the Adler zero, along the lines of Ref.~\cite{PhysRevD.65.076008}.\section{\label{sec:Adler} $\pi$-$\pi$ scattering}
\subsection{\label{sec:impulseapprox}$\pi$-$\pi$ scattering in impulse approximation}

\begin{figure*}
\includegraphics[clip=0cm,width=9cm]{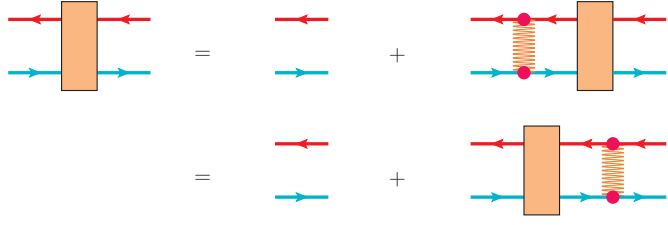}
\caption{(Color online) The self-consistent equations for the unamputated 
quark-antiquark scattering amplitude, denoted by the orange box.}
\label{fig:ladder}
\end{figure*}

We start by calculating the $\pi$-$\pi$ scattering amplitude in the impulse 
approximation, and we show that, in order to obtain the Adler zero, one has to
go beyond impulse approximation. 

The box diagram $D_O$ ($s$-channel amplitude) of the full impulse contribution
 (sum of $s$, $u$, and $t$-channel amplitudes) to $\pi$-$\pi$ scattering is depicted in Fig.~\ref{fig:box_pipiscattering}.
In the CST, it is proportional to~\cite{PhysRevD.65.076008}
\bea
D_O&\propto&-\mathrm i\int_{k0}\, \mathrm {tr} [\bar \Gamma^\pi_R (k+P_1-P_4,k+P_1) \tilde S(k+P_1)
\nonumber\\&&\,\times  \Gamma^\pi_R (k+P_1,k) \tilde S(k)  
\Gamma^\pi_R (k,k-P_2) \nonumber\\&& \times \tilde S(k-P_2) \bar 
\Gamma^\pi_R (k-P_2,k-P_2+P_3)  \nonumber\\&& \times \tilde S(k+P_1-P_4)]\,,
\eea
where $\Gamma^\pi_R$ is the reduced pion vertex function.
In the chiral limit and in the pion rest frames ($P^\mu_i=0$ where $i=1,2$ label the two incoming and $i=3,4$ the two outgoing pions) the pion vertex 
functions are given by Eq.~(\ref{eq:Gammachi}), and therefore $D_{O}$ becomes 
\bea
D_{O\chi}&\propto&-\mathrm i\int_{k0}\, \frac{A_\chi^4}{(1-B_\chi)^4(M_\chi^2-k^2)^4}
\nonumber\\ 
&&\times\mathrm {tr} [\gamma^5(M_\chi+\slashed{k})\gamma^5(M_\chi+\slashed{k})
\gamma^0\gamma^{5\dag}\gamma^0 \nonumber\\ &&\times(M_\chi+\slashed{k}) 
\gamma^0\gamma^{5\dag}\gamma^0(M_\chi+\slashed{k})]
\nonumber\\
&=&-\mathrm i \int_{k0} \frac{A_\chi^4}{(1-B_\chi)^4(M_\chi^2-k^2-
\mathrm i\epsilon)^2}\,.
\eea
 This integral has 2 double poles at 
$k_0=\pm \sqrt{m_\chi^2+{\bf k}^2}\mp \mathrm i\epsilon\linebreak=\pm E_k\mp \mathrm i\epsilon$.  Introducing the energy of the running mass, ${\cal E}_k=\sqrt{M_\chi^2+{\bf k}^2}$,
and  retaining only the 
residues of the propagator pole contributions, one obtains
\bea
 D_{O\chi}&\propto&-\mathrm i \int_{k0} \, \frac{M^4_\chi} {(k_0-{\cal E}_k+
\mathrm i\epsilon)^2 (k_0+{\cal E}_k-\mathrm i\epsilon)^2}
\nonumber\\&=& 
  \frac 12 2\pi \int_k\Bigg\{\left[ \frac{4m_\chi^3 M'_{\chi 0}}{2 E_k}-
\frac{m_\chi^4}{4 E_k^3} \left(1+2m_{\chi} M'_{\chi 0}\right)\right] 
\nonumber\\&&-
\left[-\frac{4m_\chi^3 M'_{\chi 0}}{2 E_k}+\frac{m_\chi^4}{4 E_k^3} \left(1-2m_{\chi} M'_{\chi 0}\right) \right] \Bigg\}
\nonumber\\
&=& \pi\int_k
\left[ \frac{4m_\chi^3 M'_{\chi 0}}{E_k}-\frac{m_\chi^4}{2 E_k^3} \right] \neq 0\,,
\label{eq:DOchi}
\eea 
where 
\bea
M'_{\chi 0}&\equiv& \frac{\mathrm d M_\chi(k^2)}{\mathrm d k^2}\big|_{k^2=m_\chi^2}
=\frac1{2E_k}\frac{\mathrm d M_\chi(k^2)}{\mathrm d k_0}\big|_{k^2=m_\chi^2}\, .
\eea 
The two terms in (\ref{eq:DOchi}) are non-zero, and they do not cancel. The same result is obtained for $D_Z$ and $D_X$. 
\textit{One concludes that, in order to obtain the Adler zero from the amplitude 
in the chiral limit, one has to go beyond the impulse approximation}. 
Therefore, the calculation of the quark-quark ladder sum to include intermediate-state interactions is unavoidable for 
crucial cancellations to occur. 
To achieve this, we extend the strategy of 
Refs.~\cite{PhysRevD.65.076008,PhysRevC.67.035201} to accommodate scalar, pseudoscalar, and tensor linear confinement in the CST formalism. 

\subsection{\label{sec:Prerequisites } Prerequisites}
\subsubsection{Axial-vector Ward-Takahashi identity and the ladder sum}

Because we are going  to deal with diagrams which include a ladder sum in the intermediate state and at each vertex, it is useful to establish a Ward-Takahashi identity for the axial vertex when 
\lq\lq sandwiched'' between two ladder sums. In order to derive this identity, it is convenient to introduce some definitions and useful relations. First, 
we introduce the \lq\lq unamputated'' quark-antiquark scattering 
amplitude in the ladder approximation, $L(p_1',p_2';p_1,p_2)$.  It includes the external  propagators, two from the initial state and two from the final state (except for the inhomogeneous term, which has only two).   Using a direct product representation, with $[\tilde S(k_1) \otimes \tilde S(k_2)]\equiv \tilde S_{\alpha\alpha'}(k_1) \tilde S_{\beta'\beta}(k_2)$, where $\alpha,\alpha'(\beta,\beta')$ are the Dirac indices for particle 1(2), so that, for example,
\bea
[\tilde S(k_1)\otimes  \tilde S(k_2)] {\Gamma}(k_1,k_2)&\equiv&\tilde S_{\alpha\alpha'}(k_1) \tilde S_{\beta'\beta}(k_2)\Gamma_{\alpha'\beta'}(k_1,k_2)
\nonumber\\
&=& \big[\tilde S(k_1) \Gamma(k_1,k_2)\tilde S(k_2)\big]_{\alpha\beta}\,,\nonumber\\
\eea
the ladder sum (frequently referred to simply as the ``ladder'') is 

\begin{eqnarray}
L(p_1',&&p_2';p_1,p_2)
\nonumber\\&&= -\mathrm i\, [\tilde S(p'_1)\otimes  \tilde S(p'_2)](2\pi)^4\delta^4 (p-p')
\nonumber\\&&\;\;+ [\tilde S(p'_1)\otimes  \tilde S(p'_2)]\,
 \mathrm i \int_{k0} 
\mathcal V_R(p'-k)  L(k_1,k_2;p_1,p_2)\nonumber\\
&&=  -\mathrm i\, [\tilde S(p'_1)\otimes  \tilde S(p'_2)](2\pi)^4\delta^4 (p-p')
\nonumber\\&&\;\;+ \, \mathrm i\int_{k0} 
 L(p'_1,p'_2;k_1,k_2)\mathcal V_R(k-p)  [\tilde S(p_1) \otimes  \tilde S(p_2) ]\,,\nonumber\\\label{eq:BSLadder}
\end{eqnarray}
where $p_1^{(\prime)}=p^{(\prime)}+P^{(\prime)}/2$ and $p_2^{(\prime)}=p^{(\prime)}-P^{(\prime)}/2$, and the phases are as given in Ref.~\cite{Gross:1993zj}, with a factor of $-\mathrm i$ for each propagator, vertex function ({\it except\/} pseudoscalar or axial-vector vertices, which have no such factor), kernel, or scattering amplitude, an overall factor of $\mathrm i$, and an additional factor of $-1$ for each closed fermion loop.  
This sum is diagrammatically depicted in Fig.~\ref{fig:ladder}. 

\begin{figure*}
\includegraphics[clip=0cm,width=9cm]{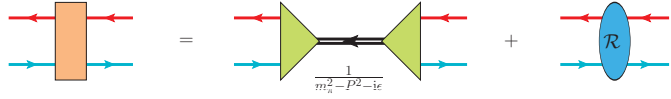}
\caption{(Color online) Spectral decomposition of the ladder with the pion pole at $m_\pi^2$ explicitly displayed}
\label{fig:SpectralDec}
\end{figure*}

It is shown in the Appendix how the insertion of the axial vertex into line 1 of an infinite ladder sum can be reduced using the Ward-Takahashi identity.  The result is   
\vspace{0.1in}
\begin{widetext}
\begin{eqnarray}
\left<L\big|\Gamma^A_R \big|L\right>&=&\int_{k0}
L(p_1',p_2';k_1',k_2)\Big[\Gamma^A_{R}(k_1',k_1)\otimes \tilde S^{-1}(k_2)\Big]
L(k_1,k_2;p_1,p_2)
\nonumber\\ 
&=&-\mathrm i\,(\gamma^5  \otimes {\bf 1}) 
L(p_1',p_2';p_1,p_2) -\mathrm i\,L(p_1',p_2';p_1,p_2) (\gamma^5  \otimes {\bf 1})\nonumber\\&&\quad
+ \,\mathrm i \int_{k0'}\int_{k0} L(p_1',p_2';k_1',k_2')\Big[{\cal V}_R(k'-k) (\gamma^5  \otimes {\bf 1})+(\gamma^5  \otimes {\bf 1}) {\cal V}_R( k'-k)\Big]
L(k_1,k_2;p_1,p_2)\,.\qquad
\label{eq:WTIL}
\end{eqnarray}
When the kernel ${\cal V}$ anticommutes with $\gamma^5$ [which is true for the vector and axial-vector pieces of the kernel in Eq.~(\ref{eq:kernel})]
the last term 
vanishes~\cite{PhysRevC.60.035209,PhysRevC.67.035201}.
The final result in this case was given in Refs.~\cite{PhysRevD.65.076008,PhysRevC.67.035201} and is depicted in Fig.~\ref{fig:WIladder}. For the more general case when $\left\lbrace\gamma^5,
{\cal V}\right\rbrace\neq0$ the result is depicted in Fig.~\ref{fig:WIladder2}, where the last four diagrams correspond to the extension of the Ward-Takahashi identity of Fig.~\ref{fig:WIladder}.  Equation~(\ref{eq:WTIL}) will be used later.

\begin{figure*}
\includegraphics[clip=0cm,width=10cm]{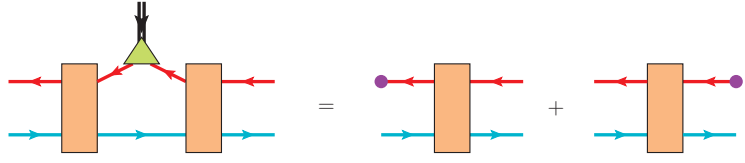}
\caption{(Color online) Ward-Takahashi identity for the ladder of a kernel with 
$\{\mathcal O_i,\gamma^5\}=0$.  The purple blobs denote $\gamma^5$'s.
}\label{fig:WIladder}
\end{figure*}

\begin{figure*}
\includegraphics[clip=0cm,width=11cm]{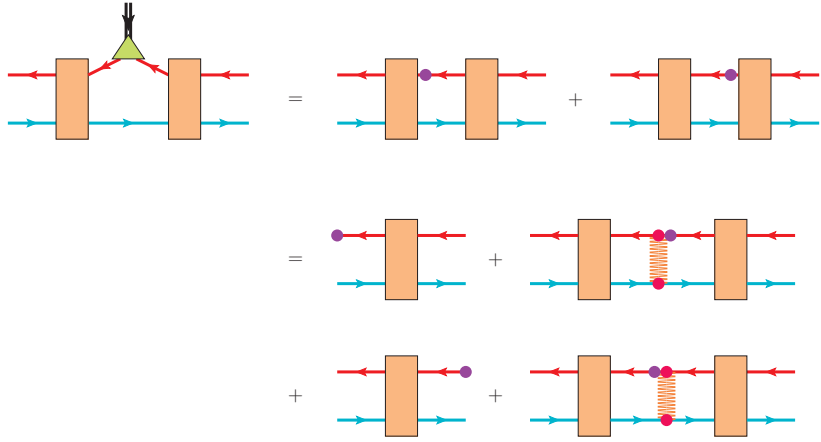}
\caption{(Color online)  By inserting the AV-WTI and ladder equation 
into the lhs of Eq. (\ref{eq:WTIL}) one obtains the Ward-Takahashi identity for the 
ladder.
}\label{fig:WIladder2}
\end{figure*}

Applying the ladder equation~(\ref{eq:BSLadder}), we can rewrite the BSE (\ref{eq:CSTBSEGammaA}) for the axial 
vertex $\Gamma^A_R$ \cite{PhysRevC.67.035201}.  Using the direct product notation,
\begin{eqnarray} 
-\mathrm i\,[\tilde S(p_1) 
\otimes \tilde S(p_2)]\Gamma^{A}_R (p_1,p_2)&=&-\mathrm i\,[\tilde S(p_1) 
\otimes \tilde S(p_2) ]\gamma^{A}_R (p_1,p_2)+\int_{k0} 
[\tilde S(p_1) 
\otimes \tilde S(p_2) ]\mathcal V_R(p-k) [\tilde S(k_1)\otimes \tilde S(k_2)] \gamma^{A}_R (k_1,k_2)  
\nonumber\\&&+ \mathrm i\int_{k0} [\tilde S(p_1) 
\otimes \tilde S(p_2) ]\mathcal V_R(p-k) [\tilde S(k_1) 
\otimes \tilde S(k_2) ] \int_{k0'}  \mathcal V_R(k-k') [\tilde S(k'_1) 
\otimes \tilde S(k'_2) ] \gamma^{A}_R (k_1',k_2') 
\nonumber\\&&+\ldots
\nonumber\\&=&\int_{k0} L(p_1,p_2;k_1,k_2) \gamma_R^{A} (k_1,k_2)\, .
\label{eq:CSTBSEGammaA2}
\end{eqnarray}

\subsubsection{Spectral decomposition of the ladder sum}

We apply the spectral decomposition of the 
ladder, assuming that it contains a bound-state pole at $P^2=m_\pi^2$, the pion pole. The ladder amplitude can then be related to the reduced bound-state vertex function
for the pion as follows:
\begin{eqnarray}
L&&(p_1',p_2';p_1,p_2)
=[\tilde S(p_1')\otimes \tilde S(p_2')]\frac{\Gamma^\pi_R(p_1',p_2')\bar 
\Gamma^\pi_R(p_2,p_1)}{m_\pi^2-P^2-\mathrm i \epsilon} 
[\tilde S(p_1)\otimes \tilde S(p_2)] 
+\mathcal R (p_1',p_2';p_1,p_2)\,, \label{eq:SpectraDeLadder}
\end{eqnarray}
where  $\mathcal R$ is the regular remainder 
 at $P^2=m_\pi^2$ which also includes the poles of all the other meson states. The only assumption we make about $\mathcal R$ is that none of its poles resides exactly at the pion mass, which is of course satisfied for any kernel that describes the meson spectrum and that is consistent with S$\chi$SB. Note that the sign of the pole term is positive because the pion is a pseudoscalar bound state (it would be negative for a scalar bound state), and that the separation between pole  
and non-pole terms is not unique away from the pole.  Equation (\ref{eq:SpectraDeLadder}) 
is shown graphically in 
Fig.~\ref{fig:SpectralDec}.
\end{widetext}
\subsubsection{Relation for the off-shell pion vertex function}

\begin{figure*}
\includegraphics[clip=0cm,width=6cm]{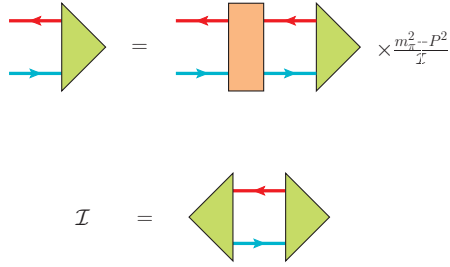}
\caption{(Color online) Relation for the off-shell pion vertex function.
}\label{fig:offshellBSEVertex}
\end{figure*}

A useful relation for inserting a ladder at a pion vertex function is obtained from Eq.~(\ref{eq:SpectraDeLadder}) by multiplying by i times the vertex function from the 
right and integrating over $p$. One obtains
\begin{eqnarray}
[\tilde S(p_1')&&\otimes \tilde S(p_2')]\Gamma^\pi_R(p_1',p_2')\nonumber\\&&\;=\frac{m_\pi^2-P^2}{\mathcal I}
\mathrm i\int_{p0} L(p_1',p_2';p_1,p_2)\Gamma^\pi_R (p_1,p_2) \nonumber\\&&\;\;-\frac{m_\pi^2-P^2}{\mathcal I}
\mathrm i \int_{p0} \mathcal R_\pi (p_1',p_2';p_1,p_2)\Gamma^\pi_R (p_1,p_2)
\, ,\qquad
\label{eq:SpectraDeLadder3a}\,,
\end{eqnarray} 
where
\begin{eqnarray}
\mathcal I&=&\mathcal I(P)
\nonumber\\
&=&\mathrm i\int_{k0} \mathrm {tr} \Big[\bar \Gamma^\pi_R (k_2,k_1) [\tilde S(k_1) 
\otimes \tilde S(k_2)]
\Gamma^\pi_R(k_1,k_2) \Big]\,\qquad
\end{eqnarray}
and $\mathcal R_\pi$ is the part of $\mathcal R$ that couples to the pion channel. Since the integral of the second term on the rhs of Eq.~(\ref{eq:SpectraDeLadder3a}) involving $\mathcal R_\pi$ has no poles at $P^2=m_\pi^2$, this term can be dropped because at the end of the calculation we will only be interested in \emph{on-shell} pion momenta $P$ for which the factor $m_\pi^2-P^2$ becomes zero. Alternatively, since the separation between the pion pole and non pole residue ${\cal R}_\pi$ is not unique away from the pion pole, we may {\it choose\/} to set ${\cal R}_\pi=0$, which {\it uniquely\/} defines the off-shell pion vertex function. We will adopt this point of view.  Without the  $\mathcal R_\pi$ term the off-shell extension of the pion vertex function is uniquely defined as the solution of      
\begin{eqnarray}
[\tilde S(p_1')&&\otimes \tilde S(p_2')]\Gamma^\pi_R(p_1',p_2')\nonumber\\&&\;=\frac{m_\pi^2-P^2}{\mathcal I}
\mathrm i\int_{p0} L(p_1',p_2';p_1,p_2)\Gamma^\pi_R (p_1,p_2) 
\, .\qquad
\label{eq:SpectraDeLadder3}\,
\end{eqnarray} 
Equation~(\ref{eq:SpectraDeLadder3}) effectively shows how one can add a ladder to---or remove it from---the pion vertex function (see Fig.~\ref{fig:offshellBSEVertex}).

\subsection{\label{sec:3meson} $\pi$-$\pi$ scattering in the chiral limit: The Adler 
self-consistency zero}

Now we are ready to calculate $\pi$-$\pi$ scattering to all orders in the chiral limit. 
Our aim is to show that, in the chiral limit, the scattering amplitude vanishes. This is known as  the Adler self-consistency zero \cite{Adler_PhysRev.137.B1022}. Our derivation closely follows the one of Ref.~\cite{PhysRevC.67.035201}.

\begin{figure*}
 \includegraphics[clip=0cm,width=11cm]{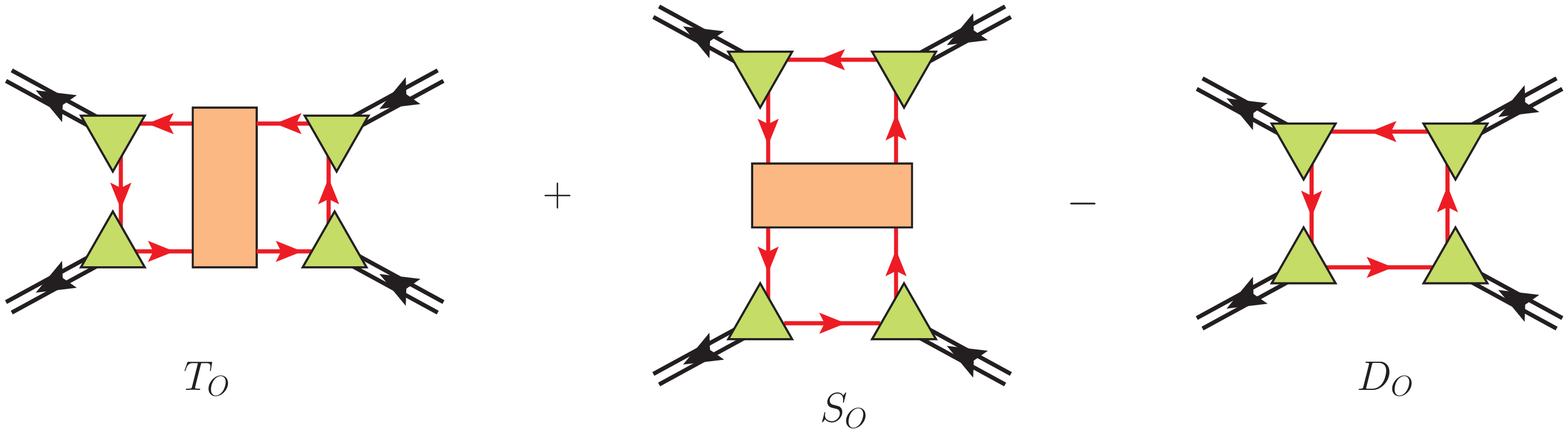}
\includegraphics[clip=0cm,width=11cm]{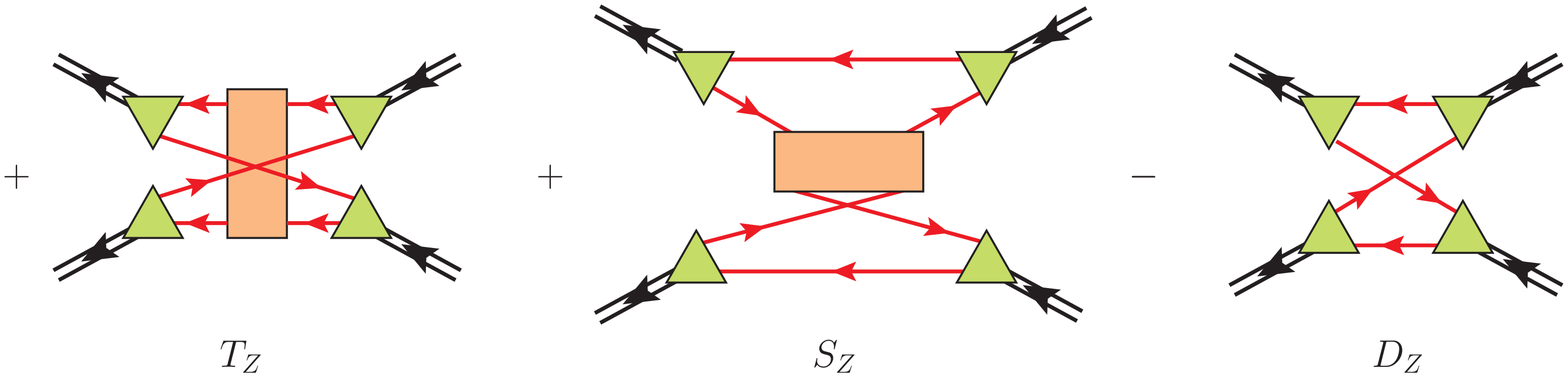}
\includegraphics[clip=0cm,width=11cm]{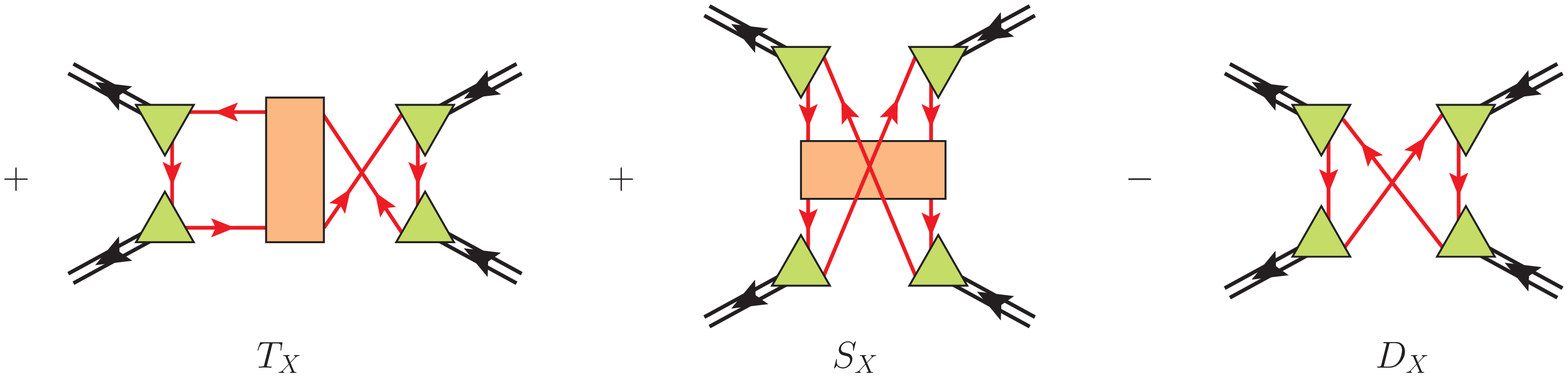}
\caption{(Color online) Contributions to $\pi$-$\pi$ scattering.
}\label{fig:pipiscattering}
\end{figure*}

There are three types of contributions, referred to as $O$, $Z$, and $X$ diagrams, which 
are shown in the three rows of Fig.~\ref{fig:pipiscattering}. 

The $D$ terms in each line, $D_O$, $D_X$, and $D_Z$, must be subtracted in order to avoid double counting of the direct contributions of 
Section~\ref{sec:impulseapprox}.
 We start our discussion by looking at the three diagrams of the first 
row (the $O$ diagrams). We will show that, in the chiral limit, the sum of the 
three diagrams vanishes, 
 \begin{eqnarray}
  T_O+S_O-D_O\longrightarrow0.\label{eq:chlOterms}
 \end{eqnarray}
 Because of the similar topologies, the sums of the diagrams in the second and third row, respectively, also vanish.
  
 We start with $T_O$.  Remembering the minus sign for a closed fermion loop,  
\begin{eqnarray}
  T_O&=& 
\int_{k0'}\int_{k0}\,\mathrm {tr} \bigg\{\Big[ \bar \Gamma^\pi_R (k'-P_3,k')\tilde S(k')\bar 
\Gamma^\pi_R (k',k'+P_4)\Big]\nonumber\\&&\times L(k'+P_4,k'-P_3;k+P_1,k-P_2)
\nonumber\\&&\times \Big[\Gamma^\pi_R(k+P_1,k) \tilde S(k)\Gamma^\pi_R(k,k-P_2)\Big]\bigg\}\,,\label{eq:TO}
\end{eqnarray}
with the ladder connecting incoming pions of momentum $P_1, P_2$ to outgoing pions with momentum $P_3, P_4$.   Note that at this stage only the remainder term ${\cal R}$ from Eq.~(\ref{eq:SpectraDeLadder}) contributes to the ladder $L$. This is because in Eq.~(\ref{eq:TO}) $L$ is projected onto two pion vertex functions and therefore its pion pole term does not contribute (there is no $\pi\to 2\pi$ coupling by G-parity conservation).

In order to evaluate this diagram, we first consider the scattering when $P_2^2\ne m_\pi^2$, and make use of the off-shell definition of the pion vertex function, 
Eq.~(\ref{eq:SpectraDeLadder3}), to insert another ladder into Eq.~(\ref{eq:TO}) by replacing $\Gamma^\pi_R(k,k-P_2)$ (this step is shown diagrammatically in the top panel of Fig.~\ref{fig:term1}). This gives 
\vspace{0.1in}
\begin{widetext}
 \begin{eqnarray}
  T_O&=&
\frac{m_\pi^2-P_2^2}{\mathcal I(P_2)}\,\mathrm i\int_{k0'}
\int_{k0}\int_{p0}\mathrm {tr}\,\bigg\{\Big[ \bar \Gamma^\pi_R (k'-P_3,k')\tilde S(k')
\bar \Gamma^\pi_R (k',k'+P_4)\Big] L(k'+P_4,k'-P_3;k+P_1,k-P_2)
\nonumber\\
&&\times\Big[\Gamma^\pi_R(k+P_1,k)\tilde S^{-1}(k-P_2) L(k,k-P_2;p,p-P_2)\Gamma^\pi_R(p,p-P_2)\Big]\bigg\}
\,.\label{eq:TOcont}
 \end{eqnarray} 
Note that in this equation, the first ladder already present in Eq.~(\ref{eq:TO}) still does not have any pion pole contribution, while the second {\it inserted} ladder contains {\it only\/} pseudoscalar contributions, including the pion pole.  Still, in anticipation of the next step, it is convenient to keep the notation general.

Next we let $P_1\to0$ (so that $P_2=P_3+P_4$), use (\ref{eq:Gammachi}) to replace $\Gamma^\pi_R(k,k)$ by $\Gamma^A_{R}(k,k)$, and then make use of the Ward-Takahashi identity (\ref{eq:WTIL}) to replace the product of the two ladders.  This generates four terms, all of which are further reduced using (\ref{eq:SpectraDeLadder}). They are depicted in the middle and bottom panels of Fig.~\ref{fig:term1} and given by

\begin{figure*}
\includegraphics[clip=0cm,width=11cm]{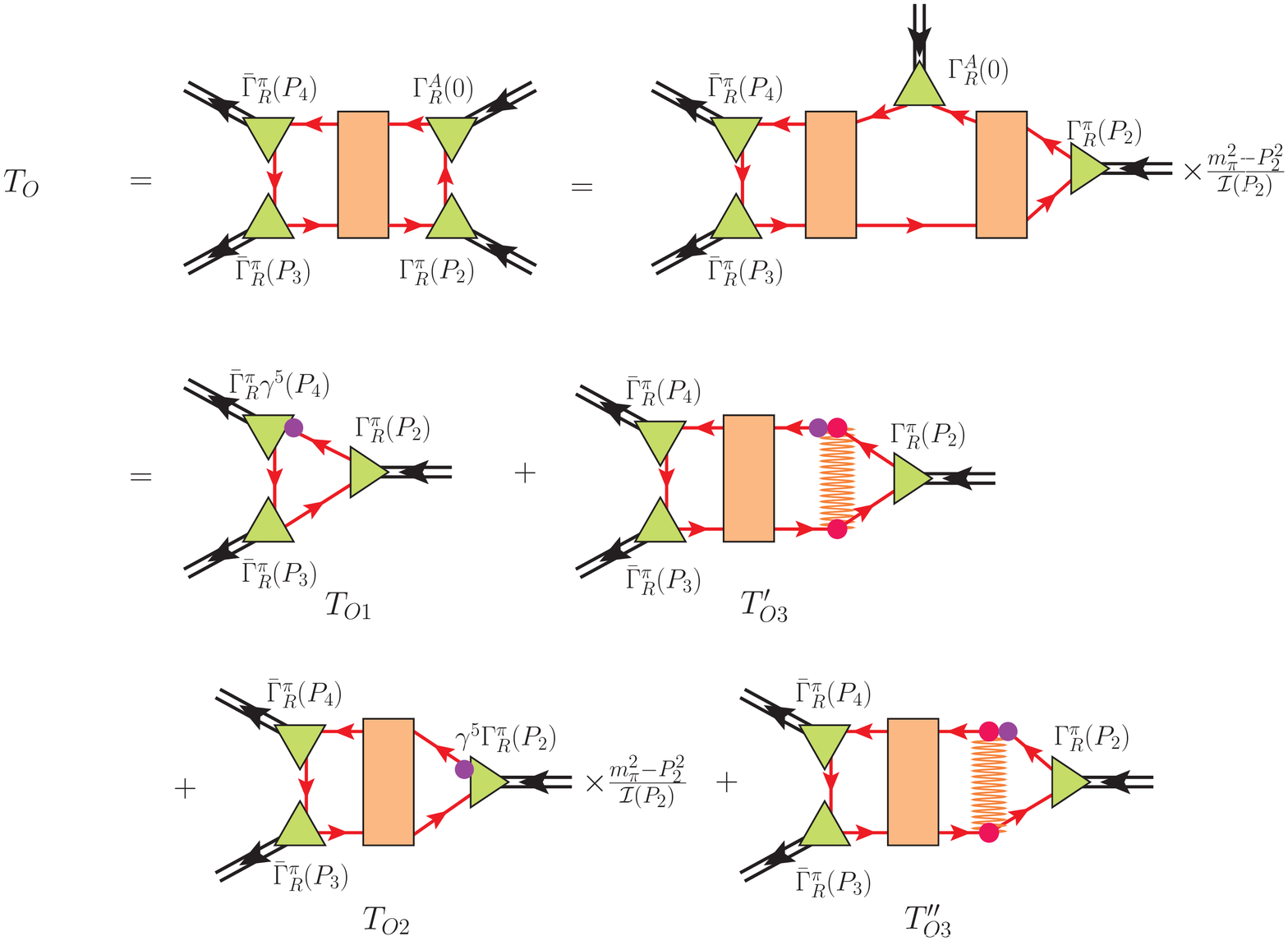}
\caption{(Color online) Expansion of $T_O$ in terms of $T_{O1}$, $T_{O2}$, $T_{O3}'$, and $T_{O3}''$. Here we use the shorthand notation $\Gamma(P)$, with $P$ being the pion momentum.  Notice the $\gamma^5$ matrix denoted by the purple blob that multiplies one of the pion vertex functions in $T_{O1}$ and $T_{O2}$.  
}\label{fig:term1}
\end{figure*}

\begin{figure*}
\includegraphics[clip=0cm,width=10cm]{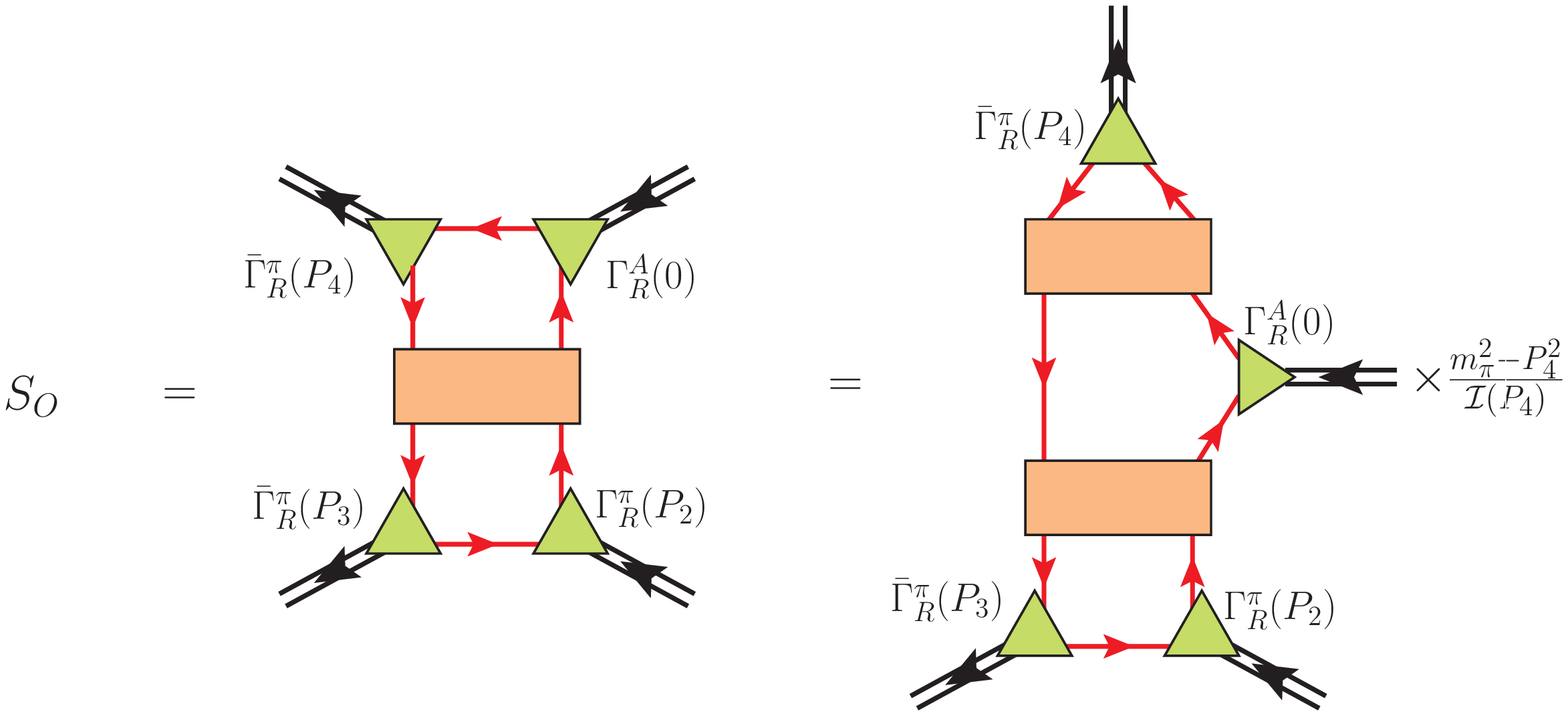}
\caption{(Color online) The first step to reduce the second term $S_O$.   
}\label{fig:term2}
\end{figure*}

 \begin{eqnarray}
  T_O&=&T_{O1}+T_{O2}+ T_{O3}' +T_{O3}''\\
 \label{eq:TO1}
T_{O1}&=&\frac{G_0 Z_0(m_\pi^2-P_2^2)}{2m_\chi\mathcal I} \int_{k0'}
\int_{p0}\mathrm {tr}\;\bigg\{ \Big[\bar \Gamma^\pi_R (k'-P_3,k')\tilde S(k')\bar \Gamma^\pi_R (k',k'+P_4)\Big] 
\gamma^5 L(k'+P_4,k'-P_3;p,p-P_2)\Gamma^\pi_R(p,p-P_2)\bigg\}
\nonumber\\
&\to&-
\frac{G_0 Z_0}{2m_\chi}\,\mathrm i
\int_{k0}\mathrm {tr}\;\Big[\bar \Gamma^\pi_R (k-P_2,k-P_4)\tilde S(k-P_4) 
\bar \Gamma^\pi_R (k-P_4,k)\gamma^5 \tilde S(k)  \Gamma^\pi_R(k,k-P_2) \tilde S(k-P_2)\Big]\,, 
\\
 T_{O2}&=& \frac{G_0 Z_0(m_\pi^2-P_2^2)}{2 m_\chi \mathcal I}\int_{k0'}
\int_{p0}\mathrm {tr}\;\bigg\{\Big[\bar \Gamma^\pi_R (k'-P_3,k')\tilde S(k')\bar \Gamma^\pi_R (k',k'+P_4)\Big] 
L(k'+P_4,k'-P_3;p,p-P_2)\gamma^5\Gamma^\pi_R(p,p-P_2) \bigg\}
\nonumber\\
&\to&-
\frac{G_0 Z_0}{2m_\chi}\frac{{\cal I}'}{{\cal I}}\,\mathrm i
\int_{k0'}\mathrm {tr}\;\Big[\bar \Gamma^\pi_R (k'-P_3,k')\tilde S(k') 
\bar \Gamma^\pi_R (k',k'+P_4) \tilde S(k'+P_4) \Gamma^\pi_R(k'+P_4,k'-P_3)\tilde S(k'-P_3)\Big]\,,
  \label{eq:TO2}
  \\
  T_{O3}&=&-\frac{G_0 Z_0(m_\pi^2-P_2^2)}{m_\chi \mathcal I}\int_{k0'}\int_{k0}\int_{p0'} \int_{p0} \mathrm {tr}\;\bigg\{
\Big[\bar \Gamma^\pi_R (k'-P_3,k')\tilde S(k')\bar \Gamma^\pi_R (k',k'+P_4)\Big] L(k'+P_4,k'-P_3;k,k-P_2)
 \nonumber\\
&&\times  V_{LR}(k-p')
 \left[\lambda_S\left(\gamma^5 \otimes \mathbf 1 +\mathbf 1\otimes \gamma^5\right)+\frac12 \lambda_T(\gamma^5\sigma^{\mu\nu}\otimes \sigma_{\mu\nu} ) \right]
 L(p',p'-P_2;p,p-P_2)\Gamma^\pi_R(p,p-P_2)\bigg\}
 \nonumber\\
 &\to&
\frac{G_0 Z_0}{m_\chi }\,\mathrm i
\int_{k0'}\int_{k0}\int_{p0'} \mathrm {tr}\;\bigg\{\Big[\bar \Gamma^\pi_R (k'-P_3,k')\tilde S(k') 
\bar \Gamma^\pi_R (k',k'+P_4)\Big]L(k'+P_4,k'-P_3;k,k-P_2) \nonumber\\
&&\times V_{LR}(k-p')
 \left[\lambda_S\left(\gamma^5 \otimes \mathbf 1 +\mathbf 1\otimes \gamma^5\right)+\frac12 \lambda_T(\gamma^5\sigma^{\mu\nu}\otimes \sigma_{\mu\nu} ) \right]
 \Big[ \tilde S(p') \otimes  \tilde S(p'-P_2)\Big]
\Gamma^\pi_R(p',p'-P_2)\bigg\}
\, ,\label{eq:TO3} 
 \end{eqnarray}
 \end{widetext}
 where $T_{O3}=T_{O3}' +T_{O3}''$, and in the second expression for $T_{O1}$ we introduced $k=k'+P_4$, and 
\begin{eqnarray}
\mathcal I'&=&\mathcal I'(P)
\nonumber\\
&=&\mathrm i\int_{k0} \mathrm {tr} \Big[\bar \Gamma^\pi_R (k_2,k_1) [\tilde S(k_1) \gamma^5
\otimes \tilde S(k_2)]
\Gamma^\pi_R(k_1,k_2) \Big].\qquad
\end{eqnarray}

Before proceeding further, it is useful to reflect on the physical content of these equations.  The first line of each equation is the result from one of the contributions from the Ward-Takahashi identity (\ref{eq:WTIL}). For example, Eq.~(\ref{eq:TO1}) collapses, symbolically,  $L_1 \Gamma^A_R L_2\to\gamma^5 L_3$ where, as already pointed out, $L_1$ [the first ladder in Eq.~(\ref{eq:TOcont})] contained no pion channel (think of a $\rho$, for example), $L_2$ [the second ladder in Eq.~(\ref{eq:TOcont})] contained the pion channel (take the $\pi$ itself), and $L_3$ [the ladder in Eq.~(\ref{eq:TO1})] is general and could contain the pion pole.  
Physically, this contribution would then represent a $\rho\pi\pi$ transition collapsing to a $\gamma^5\pi$ coupling.  
Then, the second line in each equation shows how, because of the factor $m_\pi^2-P_2^2$ from the insertion of $L_2$ multiplying the equation, only the pion pole term will survive the $P_2^2\to m_\pi^2$ limit (remember that $\mathcal R$ has no pole at $P_2^2=m_\pi^2$), reducing an initial $\rho$ contribution (in this example) in $L_1$ to a box involving three pion vertex functions and one $\gamma^5$ (it is this additional $\gamma^5$ at one pion vertex that prevents this diagram from vanishing).  This remarkable collapse of $L_1$ is a consequence of the Ward-Takahashi identity and the chiral limit. 

While $T_{O1}$ survives the chiral limit, the other terms vanish.  The term $T_{O2}\to0$ because $\Gamma^\pi_{R\chi} \propto \Gamma^A_{R\chi}$ and hence ${\cal I}'\to0$, since it is the trace of an odd number of $\gamma^5$ matrices. Physically, it is a consequence of the fact that the pion does not couple to the scalar channel.  
The reduction of $T_{O3}$, which is proportional to the anticommutator of $\mathcal O_i$ and $\gamma^5$, uses the results from Eq.~(\ref{eq:WTIL3}) which show that only contributions from the scalar, pseudoscalar, and tensor parts of the linear confining kernel will contribute. However, because of the decoupling of the linear confinement kernel from the zero-mass pion equation discussed in Sec.~\ref{sec:confinement}, each of these contributions \textit{integrates to zero in the chiral 
limit}, and therefore $T_{O3}\to0$.  The only contribution from $T_O$ to survive in the chiral limit is the triangle contribution $T_{O1}$.  

By considering a pion vertex with $P_4^2\ne m_\pi^2$ the $S_O$ diagrams can be computed in a similar way (the first step is shown in Fig.~\ref {fig:term2}). The only term to survive is $S_{O1}$, the analogue of $T_{O1}$.  Comparing Figs.~\ref{fig:term1} and \ref{fig:term2}  shows that the figures are identical if $P_4\leftrightarrow -P_2$, since $P_1=0$ and $P_3=P_2-P_4$ is unchanged in both diagrams. Starting from this observation, $S_{O1}$ can be transformed using the properties of the  charge conjugation operation on the pion vertices, and the propagators
\bea
{\cal C}\Gamma^{ \pi\intercal}_R(p_1,p_2){\cal C}^{-1}&=&\Gamma^\pi_R(-p_2,-p_1)\,,\cr
{\cal C}\tilde S^{\intercal}(p){\cal C}^{-1}&=&\tilde S(-p)\, .
\eea
This leads to

\vspace{0.1in}
\begin{widetext}
\begin{eqnarray}
S_{O1}&=&-\frac{G_0 Z_0}{2m_\chi}\,\mathrm i
\int_{k0}\mathrm {tr}\;\Big[\bar \Gamma^\pi_R (k+P_4,k+P_2)\tilde S(k+P_2) 
\bar \Gamma^\pi_R (k+P_2,k)\gamma^5 \tilde S(k)  \Gamma^\pi_R(k,k+P_4) \tilde S(k+P_4)\Big]
\nonumber\\
&=&-\frac{G_0 Z_0}{2m_\chi}\,\mathrm i
\int_{k0}\mathrm {tr}\;\Big[\Gamma^{ \pi\intercal}_R(k+P_4,k+P_2)\tilde S^{\intercal}(k+P_4)\bar \Gamma^{ \pi\intercal}_R (k,k+P_4)  \tilde S^{\intercal}(k) (\gamma^5)^{\intercal} \bar\Gamma^{ \pi\intercal}_R (k+P_2, k) \tilde S^{\intercal}(k+P_2)\Big]
\nonumber\\
&=&-\frac{G_0 Z_0}{2m_\chi}\,\mathrm i
\int_{k0}\mathrm {tr}\;\Big[\Gamma^\pi_R(-k-P_2,-k-P_4)\tilde S(-k-P_4)\bar \Gamma^\pi_R (-k-P_4, -k)  \tilde S(-k) \gamma^5 \bar\Gamma^\pi_R (-k, -k-P_2) \tilde S(-k-P_2)\Big]
\nonumber\\
&=&-\frac{G_0 Z_0}{2m_\chi}\,\mathrm i
\int_{k0}\mathrm {tr}\;\Big[\Gamma^\pi_R(k-P_2,k-P_4)\tilde S(k-P_4)\bar \Gamma^\pi_R (k-P_4,k)  \tilde S(k) \gamma^5 \bar\Gamma^\pi_R (k, k-P_2) \tilde S(k-P_2)\Big]\,,\label{eq:SO1}
  \end{eqnarray}
 where, in the last line, we changed $k\to-k$, a transformation which also holds for the $k0$ prescription discussed above.

 Next, the box diagram, for the special case when $P_1=0$, can be written 
  
  \begin{eqnarray}
  D_O&=& -\mathrm i
\int_{k0} \mathrm {tr}\; \Big[\bar \Gamma^\pi_R(k-P_2,k-P_4)\tilde S(k-P_4)
\bar \Gamma^\pi_R (k-P_4,k')\tilde S(k)\Gamma^\pi_R(k,k)  \tilde S(k)\Gamma^\pi_R(k,k-P_2) \tilde S(k-P_2)\Big]
\nonumber\\
&=&-\frac{G_0 Z_0}{2m_\chi}\,\mathrm i
\int_{k0} \mathrm {tr}\;\Big[\bar\Gamma^\pi_R(k-P_2,k-P_4)\tilde S(k-P_4)
\bar \Gamma^\pi_R (k-P_4,k)\Big(\gamma^5  \tilde S(k)+\tilde S(k)\gamma^5\Big)\Gamma^\pi_R(k,k-P_2) \tilde S(k-P_2)\Big] \, , 
\nonumber\\
&& \label{eq:DO}
  \end{eqnarray}
  \end{widetext}
 where the second line first replaces the chiral limit of $\Gamma^\pi_R(k,k)$ by $\Gamma^A_{R\chi}$ using Eq.~(\ref{eq:Gammachi}) and then uses the AV-WTI (\ref{eq:AVWTI}). From Eqs.~(\ref{eq:TO1}),~(\ref{eq:SO1}),~and~(\ref{eq:DO}) 
we find that 
$
  T_{O1}+S_{O1}-D_O=0\,,
$
which completes the proof of Eq.~(\ref{eq:chlOterms}).

Analogous considerations apply, of course, also to the $Z$ and $X$ diagrams.  
This constitutes 
the proof of the Adler self-consistency zero.

\subsection{Gell-Mann--Oakes--Renner Relation}

Although it is not directly related to $\pi$-$\pi$ scattering, the Gell-Mann--Oakes--Renner relation is an important consequence of the AV-WTI. It is interesting to determine its form in the CST framework, because it involves the quark condensate given in terms of the dressed quark propagator, while in the CST we use damped dressed propagators. 

To derive it, we extend the strategy of Ref.~\cite{PhysRevC.67.035201}. Starting with the CST-BSE for $\Gamma^A_{R}$, Eq.~(\ref{eq:CSTBSEGammaA2}), inserting the spectral 
decomposition~(\ref{eq:SpectraDeLadder}), and neglecting terms of order $m_0$ and $P$ gives
\vspace{0.1in}
\begin{widetext}
\bea
\Gamma_{R}^A (p_1,p_2)&=&
 [\tilde S(p_1)\otimes \tilde S(p_2)]^{-1}\,\mathrm i
\int_{k0} L(p_1,p_2;k_1,k_2)   
\gamma_R^{A} (k_1,k_2) 
\nonumber\\&=&
\frac{\Gamma^\pi_R (p_1,p_2)}{m_\pi^2-P^2-\mathrm i\epsilon}
\,\mathrm i \int_{k0} \,\mathrm {tr}\big[\tilde S(k_1)
\bar \Gamma^\pi_R (k_1,k_2)\tilde S(k_2) \gamma_R^{A} (k_1,k_2)\big]\,.
\label{eq:CSTBSEGammaA4}
\eea%
\end{widetext}
Taking the $P^2\to0$ limit of both sides, and using the relation (\ref{eq:Gammachi}) to cancel the common factor of $\Gamma^A_{R\chi}$ (where we neglect terms of order $m_\pi$ and $P$ in the difference between 
$\Gamma_{R\chi}^A$ and $\lim_{P^2\to0}\Gamma_R^A$), gives the condition
\bea
1=\frac{G_0 Z_0}{2m_\chi m_\pi^2}
\,\mathrm i \int_{k0} \,\mathrm {tr}\big[ \tilde S(k)
\bar \Gamma^\pi_R (k,k)\tilde S(k) \gamma_R^{A} (k,k)\big] \, .
\label{eq:CSTBSEGammaA5-1}
\eea

Next note that, in our model, the pion 
decay constant $f_\pi$ is defined by~\cite{PhysRevC.67.035201} 
\begin{eqnarray}
\sqrt{2}f_\pi P^\mu=\mathrm {i}\int_{k0}  \,\mathrm {tr}\big[ \tilde S(k)
\bar \Gamma^\pi_R (k,k)\tilde S(k) \gamma_R^{5\mu} (k,k)\big]\,.\qquad
\label{eq:fpi}
\end{eqnarray}
Contracting~(\ref{eq:fpi}) with $P_\mu$ and comparing it with Eqs.~(\ref{eq:CSTBSEGammaA5-1}) 
and~(\ref{eq:gammaA}), we conclude that 
\bea
\frac{f_\pi}{\sqrt{2}}=\frac{m_\chi}{G_0 Z_0}\, . \label{eq:fpig0}
\eea
Next, return to Eq.~(\ref{eq:CSTBSEGammaA5-1}) and use Eq.~(\ref{eq:Gammachi}) to replace $\Gamma^\pi_R$ by $\Gamma^A_{R\chi}$,  Eq.~(\ref{eq:gammaAP0}) to replace $\gamma^A_R$ by $\gamma^5$,  and (\ref{eq:fpig0}) to replace $G_0 Z_0$ by $f_\pi$, giving
\begin{eqnarray}
f_\pi^2 m_\pi^2&=&m_0\,\mathrm i \int_{k0}\,\mathrm {tr}\big[ \tilde S(k)\bar \Gamma^A_{R\chi} (k,k) \tilde S(k) \gamma^5\big]\frac{1}{h^2(k)} 
\nonumber\\
&=&-m_0\,\mathrm i \int_{k0}\,\mathrm {tr}\Big[ \tilde S(k) \Big(\tilde S^{-1}(k)\gamma^5+\gamma^5 \tilde S^{-1}(k)\Big) 
\nonumber\\
&&\qquad\times\tilde S(k) \gamma^5\Big]\frac{1}{h^2(k)} 
\nonumber\\
&=&-2 m_0\,\mathrm i  \int_{k0}\,\mathrm {tr}\, S(k) \,,
\label{eq:CSTBSEGammaA5}
\end{eqnarray}
where  we  used  the AV-WTI,
Eq.~(\ref{eq:AVWTI}), to replace $\bar \Gamma^A_{R\chi}$ and $\bar \gamma^5=-\gamma^5$. Notice that the dependence on the strong quark form factors has canceled. Since the quark condensate is
$\langle\bar qq\rangle\equiv\mathrm i\, \mathrm {tr}\int_k S(k)$, 
the Gell-Mann--Oakes--Renner relation follows:
\begin{eqnarray}f_\pi^2 m_\pi^2
=- 2m_0\,\langle\bar qq\rangle \,.
\label{eq:GORrel}
\end{eqnarray}

\section{Summary and Conclusions}\label{sec:summary}

This work describes the application of the CST to a dynamical quark model of $\pi$-$\pi$ scattering. More generally, we have found that it is possible to preserve the essential AV-WTI even in the presence of a linear confining interaction with scalar and pseudoscalar components, provided only that these components have equal weight. (No restriction is placed on the strength of any vector, axial-vector, or tensor components of the confining interaction, nor on the vector or axial-vector components of any other type of interaction.)   With a kernel with these limitations,  the AV-WTI is satisfied and we show that, as a consequence, the Adler zero in the $\pi$-$\pi$ scattering amplitude emerges automatically. This feature allows the CST model to be applied to both 
heavy and light quark systems. 

While some of these results are shared by many other models 
with vector or axial-vector kernels that anticommute with $\gamma^5$, away from the chiral limit our linear confining interaction, if it has scalar, pseudoscalar, or tensor components, will produce contributions to the $\pi$-$\pi$ scattering lengths not present in the famous Weinberg result~\cite{Weinberg:1966kf}.  We have not yet investigated how big these contributions might be--all that we know at present is that they must vanish in the chiral limit.  Comparison of predictions for these effects with experimental data, together with the contributions of the confining interaction to the meson spectrum, will constrain the strength and spin structure of the confining interaction and will be a subject for future work.

A feature of our model is that strong quark form factors are used simultaneously (i) to describe the physical 
effects of overlapping  exchange interactions that go beyond the rainbow 
approximation, and (ii) to provide a covariant regularization scheme.   

It remains to be seen whether a scalar potential in the intermediate-quark-mass range could be thought of
as a coherent superposition of vector gluons, but if that is so, the preservation of the AV-WTI requires that it must be accompanied by a pseudoscalar exchange. This is another topic for future study.

\appendix
\begin{widetext}

\section*{Appendix}\label{appendix:B}
The identity~(\ref{eq:WTIL}) is proven as follows. We apply the AV-WTI of 
Eq.~(\ref{eq:AVWTI}) (represented in Fig.~\ref{fig:AVWTI}) on the lhs of 
Eq.~(\ref{eq:WTIL}), and then use 
the self-consistent equations for the  ladder sum, Eq.~(\ref{eq:BSLadder}) represented in Fig.~\ref{fig:ladder}, to obtain four terms:
\begin{eqnarray} 
\left<L\big|\Gamma^A_R \big|L\right>&=&\int_{k0}
L(p_1',p_2';k_1',k_2)\Big[\Gamma^A_{R}(k_1',k_1)\otimes \tilde S^{-1}(k_2)\Big]
L(k_1,k_2;p_1,p_2)
\nonumber\\
&=&\int_{k0}L(p_1',p_2';k_1',k_2)\Big[\Big(\tilde S^{-1}(k'_1)\gamma^5+\gamma^5 \tilde S^{-1}(k_1)\Big)\otimes \tilde S^{-1}(k_2)\Big]
L(k_1,k_2;p_1,p_2) \nonumber\\
\nonumber\\
&=&-\mathrm i\,(\gamma^5  \otimes {\bf 1}) 
L(p_1',p_2';p_1,p_2) -\mathrm i\, L(p_1',p_2';p_1,p_2) (\gamma^5  \otimes {\bf 1})
\nonumber\\&&\quad 
+ \, \mathrm i\int_{k0'}\int_{k0} L(p_1',p_2';k_1',k_2')\Big[{\cal V}_R(k'-k) (\gamma^5  \otimes {\bf 1})+(\gamma^5  \otimes {\bf 1}) {\cal V}_R(k'-k)\Big]
L(k_1,k_2;p_1,p_2)\,.
\label{eq:WTIL2}
\end{eqnarray}
All four terms of the rhs are depicted in Fig.~\ref{fig:WIladder2}.
The two terms with the kernel (which is a sum of operators ${\cal O}_i$) are proportional to the anticommutator  
$\left\lbrace \gamma^5,\mathcal O_i\right\rbrace$.   
For vector and axial-vector spin structures, they vanish, leaving only the 
two terms of the rhs of Fig.~\ref{fig:WIladder}. For the scalar, pseudoscalar, and tensor structures of the linear 
confining part of the kernel, 
$\left\lbrace \gamma^5,\mathcal O_i\right\rbrace\neq0$, 
and therefore we must keep these terms in all calculations. Specifically, for the kernel of Eq.~(\ref{eq:kernel}),  
Eq.~(\ref{eq:WTIL2}) becomes 
\begin{eqnarray}
\left<L\big|\Gamma^A_R \big|L\right>&=&-\mathrm i\,\left(\gamma^5\otimes\mathbf 1 \right)L(p_1',p_2';p_1,p_2) -\mathrm i\, L(p_1',p_2';p_1,p_2) \left(\gamma^5\otimes\mathbf 1 \right)\nonumber\\
&&+2 \, \mathrm i
\int_{k0'} \int_{k0}
L(p_1',p_2';k_1',k_2')  
 \left[\lambda_S\left(\gamma^5 \otimes \mathbf 1 +\mathbf 1\otimes \gamma^5\right)+\frac12 \lambda_T(\gamma^5\sigma^{\mu\nu}\otimes \sigma_{\mu\nu} ) \right] 
V_{LR} (k'-k) 
L(k_1,k_2;p_1,p_2)\,.\nonumber\\
\label{eq:WTIL3}
\end{eqnarray}
Notice the factor of 2 since 
$\left\lbrace \gamma^5,\mathcal O_i\right\rbrace=2\gamma^5\mathcal O_i$ for $i=S,\; P,$ and $T$.
\end{widetext}
\begin{acknowledgements}
This work received financial support from Funda\c c\~ao para a Ci\^encia e a 
Tecnologia (FCT) under Grants No.~PTDC/FIS/113940/2009 and No. CFTP-FCT (PEst-OE/FIS/U/0777/2013). The research leading to these results has received funding from the European Community's Seventh Framework Programme FP7/2007-2013 under Grant Agreement No.\ 283286. This work was also partially supported by Jefferson Science Associates, LLC, under U.S. DOE Contract No. DE-AC05-
06OR23177. All diagrams have been drawn with JaxoDraw,  Ref.~\cite{Binosi200476}.
\linebreak
\end{acknowledgements}

\bibliographystyle{h-physrev3}
\bibliography{PapersDB.bib}

\end{document}